\documentclass[aps,prx,superscriptaddress,twocolumn,showpacs]{revtex4-1}
\usepackage[table]{xcolor}
\usepackage{pgf}
\usepackage[utf8]{inputenc}
\usepackage{amsmath}
\usepackage{braket}
\usepackage{amssymb}
\usepackage{amsmath}
\usepackage{bbold}
\usepackage{enumitem} 
\usepackage{appendix}
\usepackage{graphicx}
\usepackage{overpic}
\usepackage{mathtools}
\usepackage{commath}
\usepackage{romannum}
\usepackage{lipsum}
\usepackage{ulem}
\usepackage{epstopdf}
\usepackage[pdftex,bookmarks,colorlinks]{hyperref}
\usepackage{hyperref}

\usepackage{ulem}

\newcommand{\ud}{\mathrm{d}}
\newcommand{\Tr}{\operatorname{Tr}}
\newcommand{\nep}{\operatorname{e}}
\newcommand{\up}{\uparrow}
\newcommand{\mean}[1]{\left\langle #1\right\rangle}
\newcommand{\Real}{\Re\textrm{e}\,}

\newcommand{\opbdag}[1]{{\hat{b}^{\dagger}}_{#1}}
\newcommand{\opb}[1]{{\hat{b}^{\phantom \dagger}}_{#1}}
\newcommand{\bbeta}{\boldsymbol{\beta}}

\begin{document}
\title{Fragility to quantum fluctuations of classical Hamiltonian period doubling  }
\author{Reyhaneh Khasseh}
\affiliation{Max-Planck-Institut f\"ur Physik Komplexer Systeme, N\"othnitzer Stra{\ss}e 38, D-01187, Dresden, Germany}


\author{Angelo Russomanno}
\affiliation{Max-Planck-Institut f\"ur Physik Komplexer Systeme, N\"othnitzer Stra{\ss}e 38, D-01187, Dresden, Germany}

\author{Rosario Fazio}
\affiliation{Abdus Salam ICTP, Strada Costiera 11, I-34151 Trieste, Italy}
\affiliation{Dipartimento di Fisica, Universit{\`a} di Napoli ``Federico II'', Monte S. Angelo, I-80126 Napoli, Italy}

\begin{abstract}
We add quantum fluctuations to a classical {period-doubling Hamiltonian time crystal}, replacing the $N$ classical interacting angular momenta with quantum spins of size $l$. The full permutation symmetry of the Hamiltonian allows a mapping to a bosonic model and the application of exact diagonalization for quite large system size. {In the thermodynamic limit $N\to\infty$ the model is described by a system of Gross-Pitaevskii equations whose classical-chaos properties closely mirror the finite-$N$ quantum chaos.} For $N\to\infty$, and $l$ finite, Rabi oscillations mark the absence of persistent period doubling, which is recovered for $l\to\infty$ with Rabi-oscillation frequency tending exponentially to 0. For the chosen initial conditions, we can represent this model in terms of Pauli matrices and apply the discrete truncated Wigner approximation. For finite $l$ this approximation reproduces no Rabi oscillations but correctly predicts the absence of period doubling. {Our results show the instability of time-translation symmetry breaking in this classical system even to the smallest quantum fluctuations, because of tunneling effects.} 
\end{abstract}
	\maketitle
\section{Introduction}
The experimental discovery~\cite{zhang2017observation,choi2017observation} of Floquet time-crystals  few years after their theoretical prediction has been a real breakthrough. 
In analogy to ordinary crystals, time crystals appear as a consequence of breaking  time-translation symmetry in the system~\cite{khemani2019brief,sacha2017time,sacha2020time}. 
Time crystals were first introduced in 2012 by Frank Wilczek~\cite{wilczek2013superfluidity}. Following earlier attempts to identify systems able to display 
time-translation symmetry breaking, in 2015, a no go theorem by Watanabe and Oshikawa showed that this is not possible in the ground state or in thermal 
equilibrium~\cite{watanabe2015absence}. 

Among many possible non-equilibrium candidates, periodically periodically-driven (Floquet) systems have proven to be the most promising realization. Stimulated by 
the initial proposals~\cite{else2016floquet,khemani2016phase}, a large body of theoretical work has been performed~\cite{khemani2017defining,yao2017discrete,ho2017critical,pizzi2020time,smits2018observation,pizzi2019period,pizzi2021higher,russomanno2017floquet,federica,zhu2019dicke,lazarides2017fate,else2017prethermal,pizzi2021bistability,gong2018discrete,else2016pre}. A common ingredient to all case is the presence in the dynamics of a sufficient number of 
constraints that introduce ergodicity-breaking, thus impeding the system to reach an effective infinite temperature. 

Nearly all the attention, so far, has been devoted to quantum systems. Only few notable exceptions~\cite{yao2020classical,gambetta2019classical,heugel2019classical,malz2021seasonal} consider classical dynamics. {Especially interesting is the case of driven classical many-body Hamiltonian systems, where a long-lasting prethermal regime has been found~\cite{Emanuele,Bukov}, and period doubling (or period $n$-tupling with $n>2$) can appear in the prethermal regime~\cite{Pizzi1,Pizzi2}. All these systems eventually thermalize after a transient, and this fact relies on their chaotic dynamics.} Chaos is the generic situation for  a finite 
number of coupled classical Hamiltonian systems~\cite{lichtenberg1983regular}, but the situation can drastically change in the thermodynamic limit for long-range interacting systems~\cite{latora,lichtenberg1983regular,Firpo}.
The phenomenon of sub-harmonic generation (period-doubling) in a classical Hamiltonian driven many-body system was recently considered  in Ref.~\cite{khasseh2019many} and it 
was termed Hamiltonian synchronization (or classical Floquet time-crystals). One question is if this synchronization phenomenon is stable to fluctuations. 
In~\cite{khasseh2019many} this stability was discussed against thermal fluctuations, here we explore the stability against quantum fluctuations. 

Besides addressing the problem of stability to fluctuations, the present work aims to make a first step towards a model that has a time-crystalline 
phase both in the classical and in the quantum regime, so to understand their difference.  Most simply, we substitute the 
classical angular momenta with quantum spins of magnitude $l$ and find that, whenever $l$ is finite, the quantum fluctuations destroy the synchronized period-doubling motion. 
It is recovered only in the limit of infinite spin magnitude $l\to\infty$, when the dynamics becomes classical again.

In all the paper we focus on the case where the interactions are all-to-all and the correlations are therefore very strong. Moreover, the kicking exactly flips the spins and we take the initial state as fully polarized up. If the quantum fluctuations destroy the period doubling in this most favorable situation, they will destroy it also in case of imperfect flipping and faster decaying fluctuations. {What we find here is that adding even the smallest quantum fluctuations {($l\gg 1$ finite)}, one spoils the time-translation symmetry breaking in this model. Due to quantum tunneling, some Rabi oscillations incommensurate with the driving period add on the period-doubling oscillations. The response is no more synchronous with the driving and there is no more a persistent period-doubling response, so there is no more time crystal.}

The paper is organized as follows.
In Sec.~\ref{models:sec} we introduce a ``period-doubling order parameter'', a quantity which first vanishes at a time increasing with the system size if the system shows persistent period doubling in the thermodynamic limit. We add quantum fluctuations to the classical backbone of~\cite{khasseh2019many}. We do this in two ways, and we get two quantum models, both reducing to the classical one when $l$, a parameter we are going to describe, tends to infinity (we discuss this limit in some detail in Appendix~\ref{app:lim}). 

In the model-1 we simply substitute classical angular momenta with quantum spins of finite size $l$ and discuss it in Sec.~\ref{fir:sec}. By using a mapping to a bosonic model~\cite{federica,PhysRevB.103.224301} (Appendix~\ref{boso_map:sec}) and exact diagonalization for finite system size, we see that the period-doubling order parameter first vanishes at a time not scaling with the system size, which marks the destruction of period doubling. By studying the average level spacing ratio in Sec.~\ref{reg:quant} we find that the dynamics leading to this result is related to quantum chaos.

In Sec.~\ref{limit:sec} we perform the thermodynamic limit and show that the system is here described by a system of Gross-Pitaevskii equations. In that limit we see the period-doubling order parameter performing Rabi oscillations, so there is no period doubling. We see that the period of these oscillations diverges with the spin magnitude $l$ and in the limit of infinite spin the period doubling is recovered. This is in agreement with the fact that the quantum fluctuations disappear in this limit. 

In agreement with the finite-size quantum dynamics, the classical infinite-size Gross-Pitaevskii dynamics is chaotic, as the largest Lyapunov exponent shows (Sec.~\ref{lyapunov:sec}), but it is not fully ergodic and the Rabi oscillations can persist. Studying the amplitude and the frequency of the Rabi oscillations versus the parameter $K$ for different values of $l$, we see that the curves show a crossing point for $K\sim 1$, which corresponds to a transition from synchronized to trivial behaviour in the classical $l\to\infty$ limit. 

Rabi oscillations are related to the ones obtained in~\cite{russomanno2017floquet} for a single spin system. Coupling many of these systems with a small coupling $K$ the oscillations are still there but with a renormalized period; a large coupling on the opposite leads to the destruction of the Rabi oscillations and to small chaotic oscillations of the period-doubling order parameter. The correlations induced by the coupling are never strong enough to stabilize the period-doubling order parameter to a persistent finite value, against the quantum fluctuations. 

In Sec.~\ref{sec:sec} we study the model-2, where each classical angular momentum is substituted by an average of $2l$ Pauli matrices. We study this case by means of the discrete truncated Wigner approximation (DTWA), which we summarize in Sec.~\ref{dtwa:sec} and is known to  give good results for long-range interactions~\cite{pappalardi2018scrambling,schachenmayer2015many,khasseh2020discrete}. Also here we find find the disappearance of the period doubling (Sec.~\ref{reso:sec}): the period-doubling order parameter decays as an exponential in time and the decay time scale does not scale with $N$. We see that the decay time increases for increasing value of $l$ as a power law. So, for $l\to\infty$, where the system behaves classically, the period doubling persists for an infinite time, as expected. In Appendix~\ref{app2:sec} we discuss a different way to estimate this decay time which gives consistent results and discuss some technical aspect related to DTWA.

We remark that, for the chosen initial state, the model-2 is equivalent to the first one but the DTWA gives results in quantitative agreement only in the limit $l\to\infty$. For $l$ finite it is only correct in predicting the absence of period doubling in the limit of large $N$ but provides no Rabi oscillations.
\section{The  models} \label{models:sec}

We introduce quantum fluctuations in the model studied in~\cite{khasseh2019many}. It is a chain of $N$ coupled classical angular momenta undergoing a periodic pulsed driving. Here we will focus on the case with all-to-all interactions. These ones give rise to the strongest long-range correlations needed in order to stabilize a possible period-doubling phase. Indeed, in the classical case this model shows a phase with persistent period doubling in the thermodynamic limit, also in the all-to-all case. Adding the quantum fluctuations, we will show that the period doubling in the all-to-all interacting case disappears. This result implies the absence of period doubling also for faster decaying interactions (and smaller long-range correlations). The Hamiltonian is
\begin{align}\label{eq:cl_Hamiltonian}
{\cal H}(t)&=\sum_{i=1}^{N}\Big[-2J(m_{i}^z)^2-2h m_{i}^x\Big]    \nonumber\\
 &+\delta_{\tau}(t)\sum_{i=1}^{N}\left[\phi\, m_{i}^x-\frac{K}{2N}\sum_{j\neq i}{m_{i}^xm_{j}^x}\right]
\end{align}
where $\delta_{\tau}\equiv\sum_{n}\delta(t-n\tau)$~\cite{chiri_vov} and we put a factor $N$ in the denominator in order to ensure extensivity. The $m_{j}^\alpha$, $\alpha=x,\,y,\,z$ are the components of classical angular momenta which obey the angular-momentum Poisson brackets $\left\{m_i^\mu,\,m_j^\nu\right\}=\epsilon^{\mu\,\nu\,\rho}\delta_{i\,j}m_j^\rho$ where $\epsilon^{\mu\,\nu\,\rho}$  is the Ricci fully antisymmetric tensor. For $K$ and $h$ small enough and $\phi$ in a neighborhood of $\pi/2$ this classical Hamiltonian model shows a persistent period-doubling behaviour~\cite{khasseh2019many}.

{This classical model is such that when $K=0$ it is equivalent to a single degree of freedom showing entrainment with the driving, that's to say it shows a response synchronized with the one of the driving, with a period doubled respect to the driving~\cite{russomanno2017floquet}. When $K\neq 0$ and $N$ is finite, this response dies after a transient. For a region in the parameter space, the duration of this transient diverges with the system size going to infinity~\cite{khasseh2019many}. So, for $N\to\infty$, the system shows persistent collective oscillations with a period double with respect to the driving, in which all the spins behave in a synchronous way. This is a form of period-doubling time crystal, as we discuss in Sec.~\ref{Time-crist:sec}.}

In order to add quantum fluctuations to this model we can quantize the angular-momentum variables replacing them with quantum spins. The resulting Hamiltonian is
\begin{align} \label{Ham:eqn}
  \hat{H}^{(1)}(t) &= \sum_{j=1}^N\left[-\frac{J}{l}(\hat{s}_j^z)^2-2h\hat{s}_j^x\right]\nonumber\\
  &+\delta_\tau(t)\left[\phi\sum_{j=1}^N\hat{s}_i^x-\frac{K}{2Nl}\sum_{i,j=1}^N\hat{s}_i^x\hat{s}_j^x\right]\,,
\end{align}
where $\hat{s}_j^\alpha$, $\alpha=x,\,y,\,z$ are {quantum spins of magnitude} $l$ ($\hat{s}_j^2=l(l+1)$) obeying the commutation rules $[\hat{s}^\mu,\hat{s}^\nu]=i\epsilon^{\mu\nu\rho}\hat{s}^\rho$. We call $\hat{H}^{(1)}(t)$ as the model-1. Another possibility, which should give results physically similar to the first one, is performing the following substitution
\begin{equation}\label{substo:eqn}
 m_j^\alpha\to\hat{m}_j^\alpha\equiv\frac{1}{4l}\sum_{m=1}^{2l}\hat{\sigma}_{j,\,m}^\alpha\,.
\end{equation}
So we replace the classical angular momenta in Eq.~\eqref{eq:cl_Hamiltonian} with an average of $2l$ Pauli matrices, and then we multiply by $2l$. We call the resulting 
\begin{equation}\label{eq:Hamiltonian_qauntum}
\begin{split}
&\hat{H}^{(2)}(t)=\sum_{i=1}^{N}\Big[-\frac{J}{4l}\sum_{m,m'=1}^{2l}\hat{\sigma}_{i,\,m}^z\hat{\sigma}_{i,\,m'}^z-h\sum_{m=1}^{2l}\hat{\sigma}_{i,\,m}^x\Big] \\
&+\delta_{\tau}(t)\Big[\frac{\phi}{2}\sum_{m=1}^{2l}\hat{\sigma}_{i,\,m}^x-\frac{K}{16N\,l}\sum_{i,j\neq i}\,\sum_{m,m'=1}^{2l}\hat{\sigma}_{i,\,m}^x\hat{\sigma}_{j,\,m'}^x\Big]\,.
\end{split}  
\end{equation}
as the model-2. The parameter $l$ has the same symbol here and in the model-1 on purpose. Indeed, also $\hat{m}_j^\alpha$ are spin variables and because we choose as initial state the one fully polarized up (see Eq.~\eqref{inuno:eqn}), these are spins of size $l$, as well known from the rules of addition of angular momenta~\cite{Picasso:book}. So, with our initialization, the variables $\hat{s}_j^\alpha$ of the model-1 and the variables $\hat{m}_j^\alpha$ of the model-2 are exactly equivalent. In some sense, the model-2 is a spin-$1/2$ representation of the first one, amenable to be described by means of DTWA.

For any finite $l$ there are quantum fluctuations around the classical backbone Eq.~\eqref{eq:cl_Hamiltonian}. 
When $l\to\infty$ the fluctuations become irrelevant and both the models tend to become classical.
This can be seen, for instance, by using exactly the same methods discussed for the Lipkin-Meshkov-Glick model in~\cite{PhysRevB.86.184303,Sciolla_2}). For completeness, we give a sketch of this analysis in Appendix~\ref{app:lim}.

In the rest of the paper we numerically study the two models. We study the model-1 in Sec.~\ref{fir:sec} and the model-2 in Sec.~\ref{sec:sec}. In both cases we will consider the stroboscopic dynamics, that's to say we will focus on times which are an integer number of periods $t=n\tau$. More precisely, we will chose the time $n\tau$ as the time immediately before the $n$-th kick. We will show that whenever there are quantum fluctuations -- that's to say for any finite $l$ -- there is no period-doubling phase and in the limit $l\to\infty$ one recovers the period doubling, consistently with the attaining of the classical limit.
\section{Period doubling and time-crystal behaviour} \label{Time-crist:sec}
{In order to make our paper self contained, we briefly recap the main ideas about time-crystal behaviour, which appears as a period doubling in the classical limit of our model.
Time-crystal behaviour is a synonym for time-translation symmetry breaking: a driven system in the thermodynamic limit shows a response with a frequency multiple with the driving one. Thereby the discrete time translation symmetry of the driving is broken. In order  to spot time-translation symmetry breaking -- or its absence -- it is very important to 
define precise criteria which are able to distinguish this complex collective phenomenon from analogous single particle effects. Summarizing the discussion of {Refs.~\cite{else2016floquet,khemani2016phase,khemani2017defining}} -- where the relevant criteria and conditions to have a Floquet time crystal were introduced -- we can state that there must exist
an observable $\widehat{O}$ and a class of initial states $\ket{\psi}$ such that, considering 
 stroboscopic times $t=n\tau$, {the expectation value in the thermodynamic limit ($N\to\infty$)}
\begin{equation} \label{thermol:eqn}
  f(t) = \lim_{N\to\infty}\bra{\psi(t)}\widehat{O}\ket{\psi(t)}
\end{equation}
satisfies all of the three conditions
\begin{itemize}
  \item[{\em I)}] Time-translation symmetry breaking: $f(t+\tau)\neq f(t)$ while $\widehat{H}(t+\tau)=\widehat{H}(t)$.
  \item[{\em II)}] Rigidity: $f(t)$ shows a fixed oscillation period $\tau_B$ (for instance $\tau_B=2\tau$, the so-called ``period doubling'') without fine-tuned Hamiltonian parameters.
  \item[{\em III)}] Persistence: the non-trivial oscillation with fixed period $\tau_B$ must persist for infinitely long time, when the {thermodynamic limit $N\to\infty$ in Eq.~\eqref{thermol:eqn} has been performed.} 
%
\end{itemize}
%
%
We will focus here on period doubling, $\tau_B=2\tau$. In summary  we seek for a quantity -- called ``order parameter'' in analogy with standard symmetry breaking -- such that it oscillates with frequency $2\tau$ for an infinite time in the thermodynamic limit (when the size of the system $N$ tends to infinity). In our model (model-1 and model-2 are essentially equivalent) there are some limits where such a quantity can be found. }

{For instance, in the limit $K\to 0$, our model reduces to the kicked Lipkin-Meshkov-Glick model of~\cite{russomanno2017floquet} and the order parameter is provided by $s_N(t)\equiv\lim_{l\to\infty}\frac{1}{Nl}\sum_j\braket{\psi(t)|\hat{s}_j^z|\psi(t)}$. Here the role of the system size is played by $l$ which measures the number of interacting $ \hat{\boldsymbol{\sigma}}_{j,\,m'}$ spins which compose the $\hat{\bf s}_j^z$ in the model-2 representation.}

{Another interesting limit is the $l\to\infty$ limit (with $K\neq 0$). In this limit the model is classical (see Appendix~\ref{app:lim}) and can show persisting period doubling in the thermodynamic limit (in this case $N\to\infty$)~\cite{khasseh2019many}. In this case the order parameter is $s(t)\equiv\lim_{N\to\infty}\lim_{l\to\infty}\frac{1}{Nl}\sum_j\braket{\psi(t)|\hat{s}_j^z|\psi(t)}$.}

{Taking $l$ finite, it is quite natural that, if there were period doubling, it would appear in the finite-$l$ version of $s(t)$, namely
$$
  s_l(t)\equiv\lim_{N\to\infty}\frac{1}{Nl}\sum_j\braket{\psi(t)|\hat{s}_j^z|\psi(t)}\,.
$$
In order to see if this quantity shows persisting oscillations with period $2\tau$ (period doubling), we focus on its finite-$N$ version and perform a finite-size scaling in $N$. We focus therefore on
\begin{equation}\label{perdo:eqn}
  \mathcal{O}(t)\equiv(-1)^{t/\tau}\braket{\psi(t)|\hat{S}^z|\psi(t)}/N\,,
\end{equation}
where $\hat{S}^z=\sum_{j=1}^N\hat{s}_j^z$. 
We put the multiplying factor $(-1)^{t/\tau}$, because a period-doubling is expected to imply a change of sign $\braket{\psi(t)|\hat{S}^z|\psi(t)}$  at every period~\cite{khasseh2019many,russomanno2017floquet}). Thanks to the multiplying factor, the period doubling would appear as a never-vanishing value of $\mathcal{O}(t)$, which is easier to study. }

{In order to probe if there is persistent period doubling in the thermodynamic limit, one should check the presence of the following finite-size scaling: If $\mathcal{O}(t)$ first vanishes after a time $t^*$ scaling with $N$ towards infinity, then one has period doubling~\cite{khasseh2019many}. So, in the thermodynamic limit $\mathcal{O}(t)$ never vanishes and there is persistent period doubling. In the rest of the paper, we call for conciseness $\mathcal{O}(t)$ the ``period-doubling order parameter'', even if in the light of the discussion above this is a slight abuse of terminology.}

{For $l$ finite, we will see that $t^*$ never scales with the system size, implying the absence of persistent period doubling and time-crystal behavior.}
\section{Analysis of model-1} \label{fir:sec}
The Hamiltonian is given in Eq.~\eqref{Ham:eqn}. In order to probe the existence of a possible persistent period doubling, we initialize the system in the state 
$$
  \ket{\psi(0)}=\ket{l,\ldots,\,l}
$$
where all the spins are in an eigenstate of the corresponding $\hat{s}_j^z$ with eigenvalue $l$. This is the most favorable condition for the appearance of a persisting period doubling.

\begin{figure*}
 \begin{center}
  \begin{tabular}{cc}
    \begin{overpic}[width=8cm]{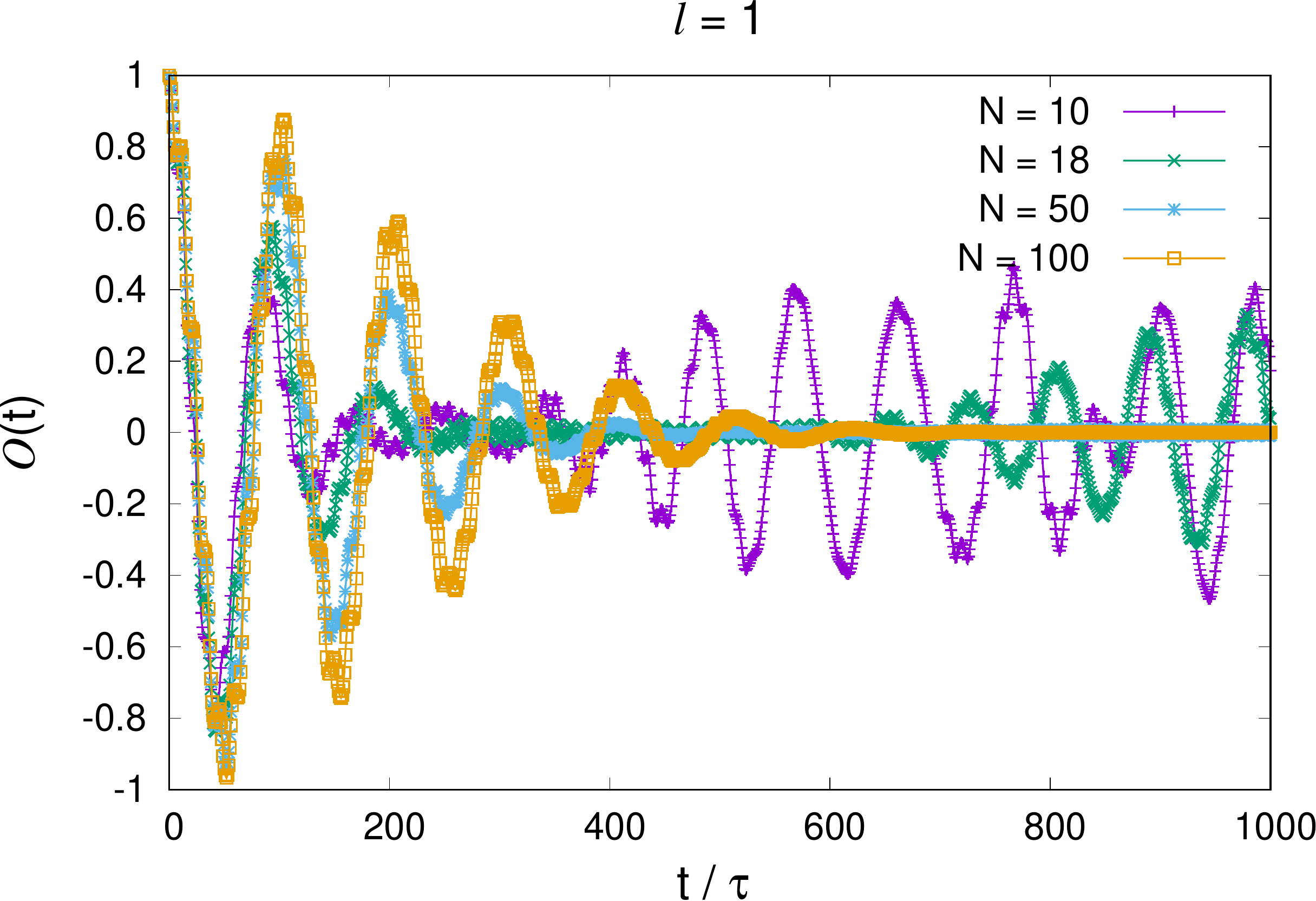}\put(30,57){(a)}\end{overpic}&
    \begin{overpic}[width=8cm]{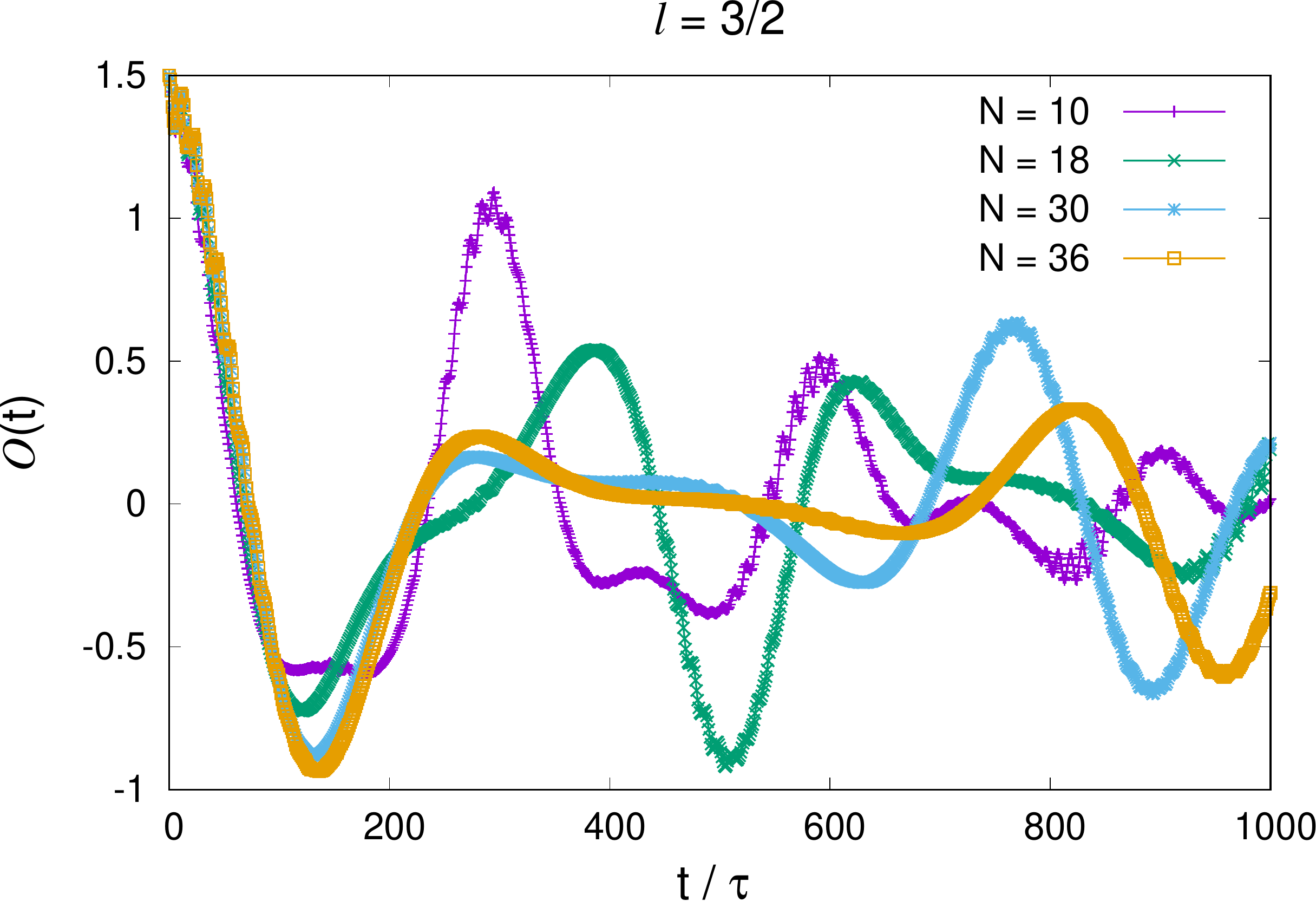}\put(20,57){(b)}\end{overpic}\\
    \begin{overpic}[width=8cm]{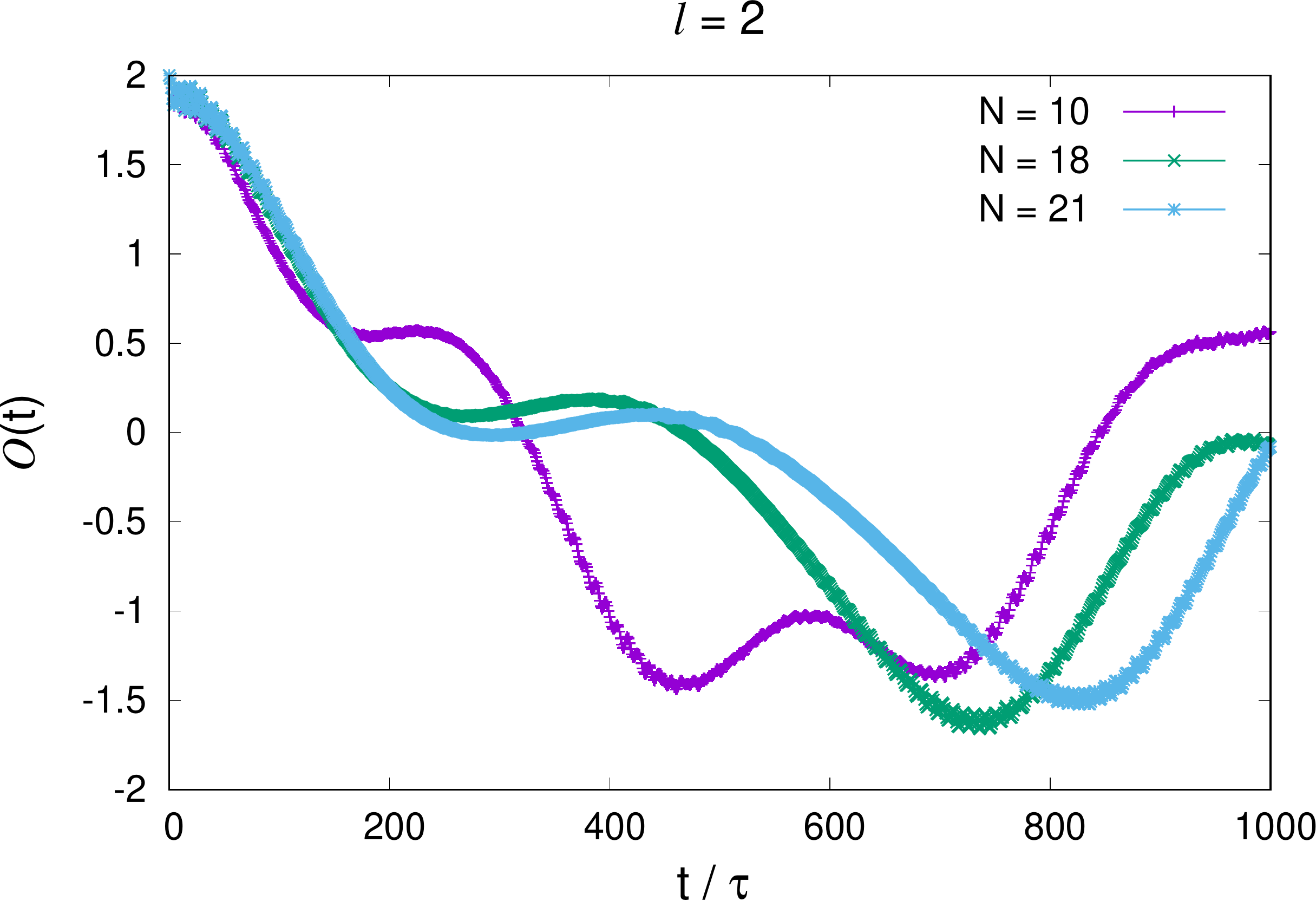}\put(20,57){(c)}\end{overpic}&
    \begin{overpic}[width=7cm]{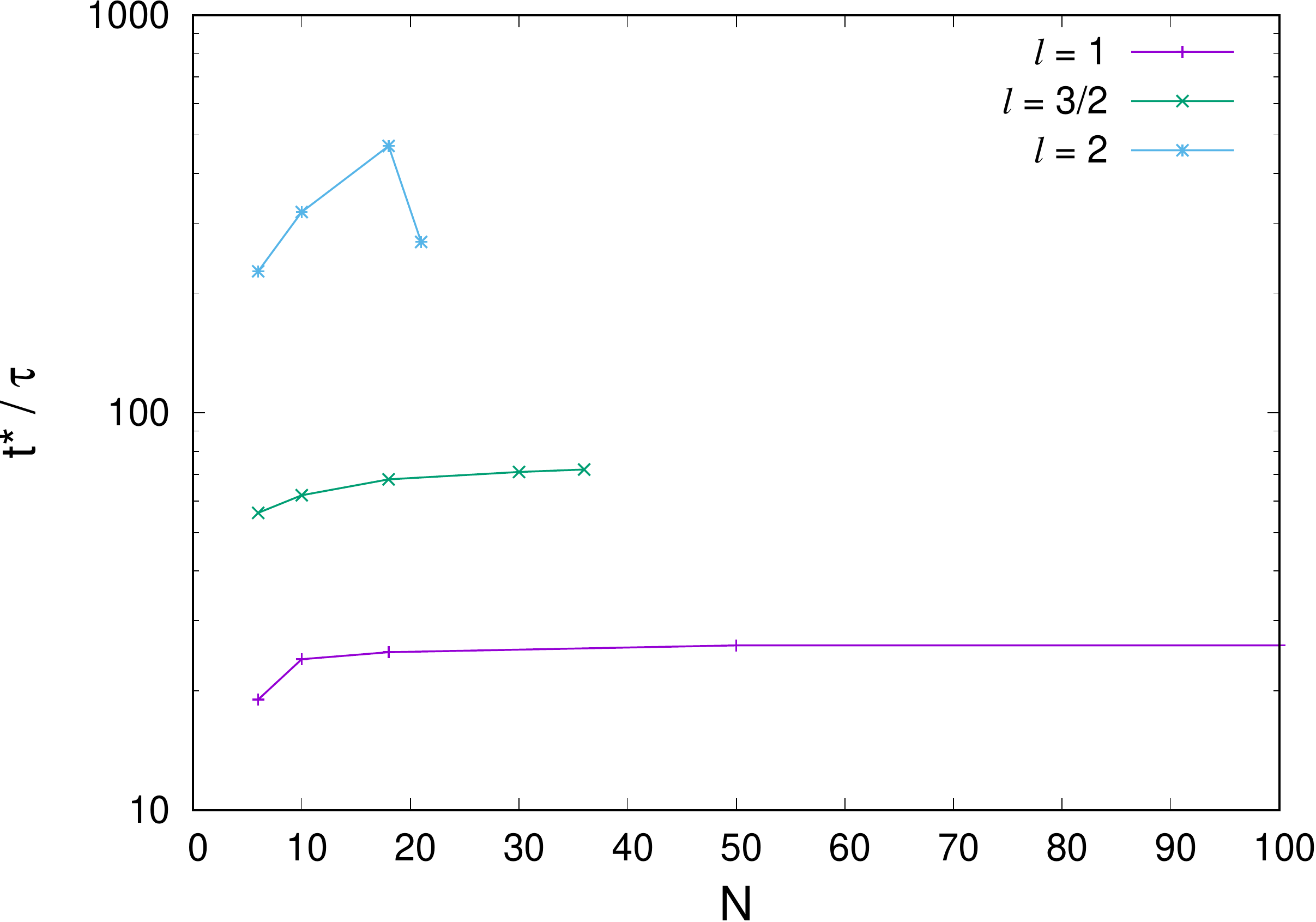}\put(80,40){(d)}\end{overpic}
  \end{tabular}
 \end{center}
 \caption{(Panels a-c) Evolution of $\mathcal{O}(t)$ with the Hamiltonian Eq.~\eqref{Ham:eqn} for different values of $l$ and $N$. We take a set of parameters giving rise to period doubling in the classical Eq.~\eqref{eq:cl_Hamiltonian} for $N\to\infty$. For finite $l$ we do not see any period-doubling behaviour in the limit of large $N$. (Panel d) Time of the first zero-crossing $t^*/\tau$ versus $N$. Numerical parameters: $h=0.1,\,\tau=0.6,\,J=1.0,\,K = 0.3,\,\phi=\pi$.}
    \label{evo:fig}
\end{figure*}

We perform the explicit derivation of the mapping in Appendix~\ref{boso_map:sec} and we find the effective bosonic Hamiltonian to be
\begin{widetext}
\begin{align} \label{ham:eqn}
  &\hat{H}(t)=-\frac{J}{l}\sum_{m=-l}^lm^2\,\hat{n}_m - h \sum_{m=-l}^{l-1}\sqrt{l(l+1)-m(m+1)}\left(\hat{b}_m^\dagger\,\hat{b}_{m+1} + \text{H.~c.}\right)\nonumber\\
    &+\delta_\tau(t)\Bigg[\frac{\phi}{2}\sum_{m=-l}^{l-1}\sqrt{l(l+1)-m(m+1)}\left(\hat{b}_m^\dagger\,\hat{b}_{m+1} + \text{H.~c.}\right)
    -\frac{K}{8Nl}\left(\sum_{m=-l}^{l-1}\sqrt{l(l+1)-m(m+1)}\left(\hat{b}_m^\dagger\,\hat{b}_{m+1} + \text{H.~c.}\right)\right)^2\Bigg]
\end{align}
\end{widetext}
with the constraint $\sum_{m=-l}^l\hat{n}_m=N$  and
\begin{equation} \label{sezz:eqn}
  \hat{S}^z = \sum_{m=-l}^lm\,\hat{n}_m\,.
\end{equation}
In the bosonic representation the initial state has the form $\ket{\psi(0)}=\frac{1}{\sqrt{N!}}(\opbdag{m})^N\ket{0}$. It is very important to remark that here the bosons jump on a linear chain of length $2l+1$, while in the clock model they used to jump over a ring. This difference in topology makes impossible the realization of the period $n$-tupling of~\cite{federica} using spin variables.
%
We choose parameters where the classical model Eq.~\eqref{eq:cl_Hamiltonian} shows period doubling and we study its fate for finite $l$ in Fig.~\ref{evo:fig}(a-c). Here we plot some examples of stroboscopic evolution of $\mathcal{O}(t)$ versus $t/\tau$ with $t=n\tau$.

We see that, fixing $l$, $\mathcal{O}(t)$ oscillates. Especially interesting is the stroboscopic time $t^*$ when $\mathcal{O}(t)$ crosses 0 for the first time. If this time increases with the system size $N$, the period-doubling oscillations persist in the thermodynamic limit and there is a period doubling. If this time saturates with $N$, the period-doubling oscillations are a transient phenomenon and there is no period doubling. We plot $t^*$ versus $N$ for the values of $l$ we have considered in Fig.~\ref{evo:fig}(d). For $l=1$ and $l=3/2$, $t^*$ saturates quite clearly with $N$. For $l=2$ there is a sudden drop and also here there is no period doubling. 
We see from Fig.~\ref{evo:fig}(d) that $t^*$ increases with $l$. This is entirely consistent with the fact that for $l\to\infty$ the model tends to the classical limit of~\cite{khasseh2019many} where there is a period doubling and $\mathcal{O}(t)$ persists indefinitely for $N\to\infty$.
%
%
\subsection{Quantum chaos}\label{reg:quant}
\begin{figure*}
 \begin{center}
  \begin{tabular}{cc}
    \includegraphics[width=8cm]{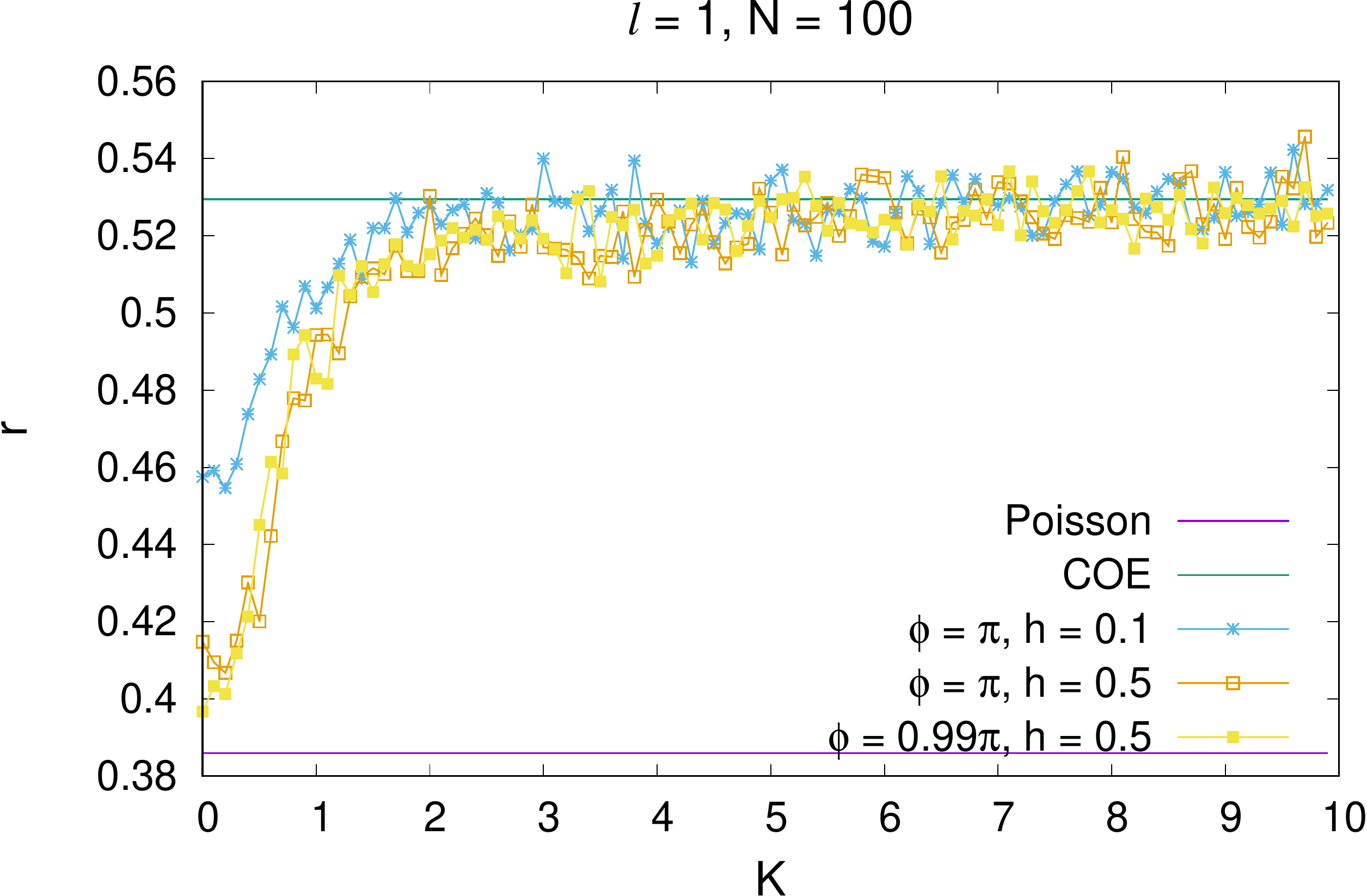}&
    \includegraphics[width=8cm]{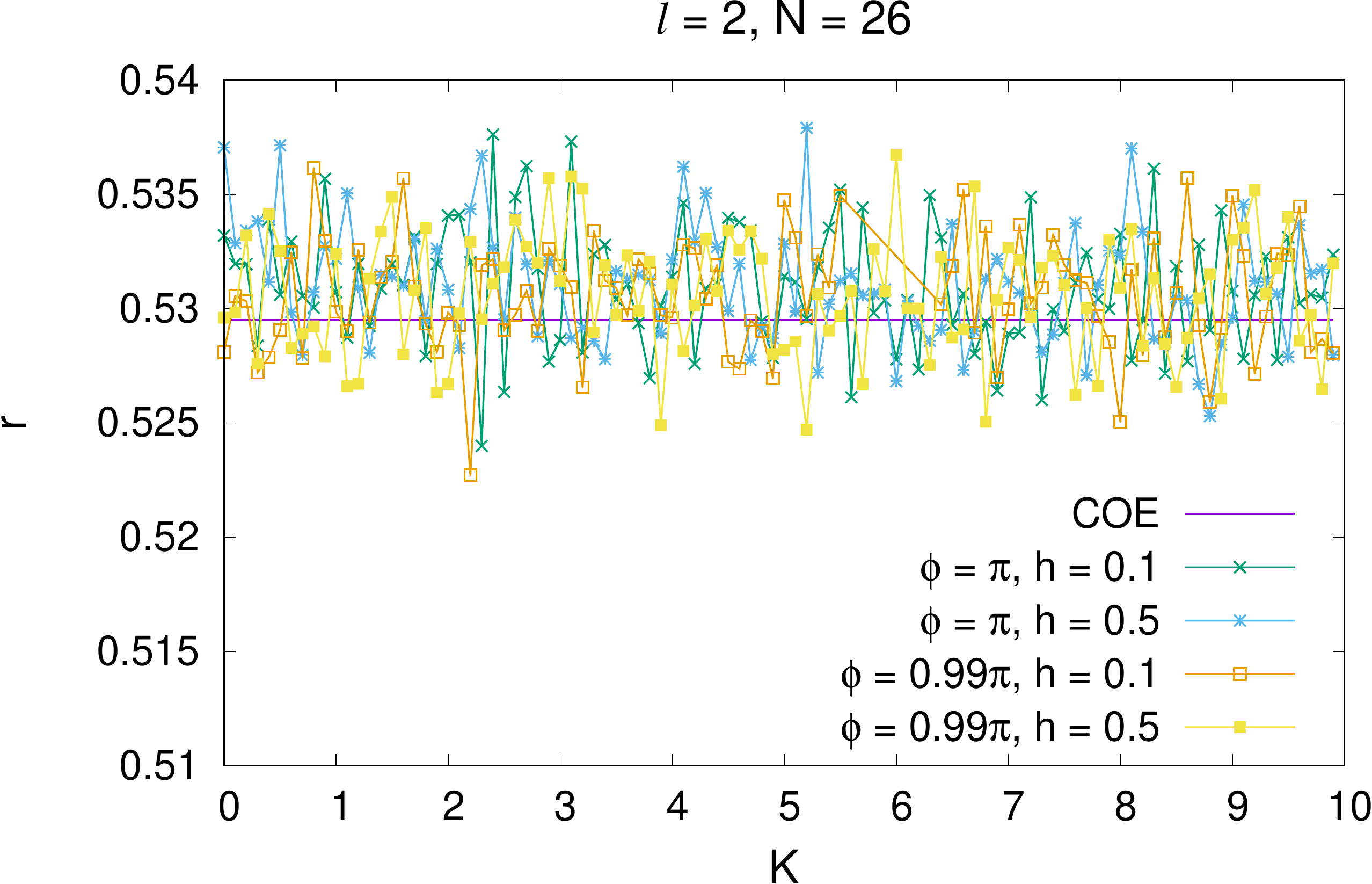}
  \end{tabular}
 \end{center}
 \caption{Average level spacing ratio $r$ versus $K$ for different values of the parameters. $J=1$.}
    \label{rorro:fig}
\end{figure*}

We can study if this dynamics is regular or quantum chaotic. ``Regular'' means similar to an integrable model where the (classical or quantum) dynamics is constrained by as many local and commuting integrals of motion as degrees of freedom~\cite{Arnold:book,essler2016quench,essler2005one}. ``Quantum chaotic'' means that the Hamiltonian is equivalent to random matrix and this leads in general to thermalization of local observables~\cite{Haake,polkovnikov2011colloquium,ue_da20}. 
In order to probe the regular or quantum-chaotic behavior, we use the average level spacing ratio, defined as~\cite{Pal_PhysRevB10}
\begin{equation}
  r \equiv \frac{1}{\dim\mathcal{H}-2}\sum_{\alpha=1}^{\dim\mathcal{H}-2}\frac{\min({\mu_{\alpha+1}-\mu_\alpha,\mu_{\alpha+2}-\mu_{\alpha+1}})}{\max({\mu_{\alpha+1}-\mu_\alpha,\mu_{\alpha+2}-\mu_{\alpha+1})}}
\end{equation}
where $\mu_\alpha$ are the Floquet levels~\cite{shirley1965solution} and $\mathcal{H}$ is the relevant Hilbert subspace (more details below). The $\mu_\alpha$ are obtained from the eigenstates $\nep^{-i\mu_\alpha\tau}$ of the time-evolution operator over one period $\hat{U}(\tau,0)$ of the Hamiltonian Eq.~\eqref{Ham:eqn} and they are taken in increasing order~\cite{nota1}. If $r\simeq 0.5269$ the level-spacing distribution is of the COE type and the dynamics is ergodic (the Floquet states are like eigenstates of a random matrix) while if $r\simeq 0.386$ the level-spacing distribution is of the Poisson type and the model is integrable (see for instance~\cite{Notarnicola_2020}). We can evaluate $r$ for the Hamiltonian in Eq.~\eqref{ham:eqn} provided we restrict to $\mathcal{H}$, the subspace even under the mirror symmetry $m\to-m$, which is an irreducible eigenspace of $\hat{U}(\tau,0)$~\cite{Berry_Les_Houches}. {We can see that $r$ reaches the quantum-chaotic COE value for $l=1$ and $K\geq 2$, while for $l=2$ the system shows always quantum chaos (see Fig.~\ref{rorro:fig}). This closely mirrors the classical-chaotic behaviour of the corresponding $N\to\infty$ Gross-Pitaevskii equations observed through the Lyapunov exponent (see Sec.~\ref{lyapunov:sec}).}

In the next subsection we consider the limit $N\to\infty$ and show that the model is described there by a system of Gross-Pitaevskii equations. In this case we will see persisting oscillations for $\mathcal{O}(t)$, for any $l$, and we will argue that they are Rabi oscillations between the states with angular momenta $Nl$ and $-Nl$.


%
\subsection{Gross-Pitaevskii equations in the $N\to\infty$ limit} \label{limit:sec}
\begin{figure*}
 \begin{center}
  \begin{tabular}{cc}
    \begin{overpic}[width=8cm]{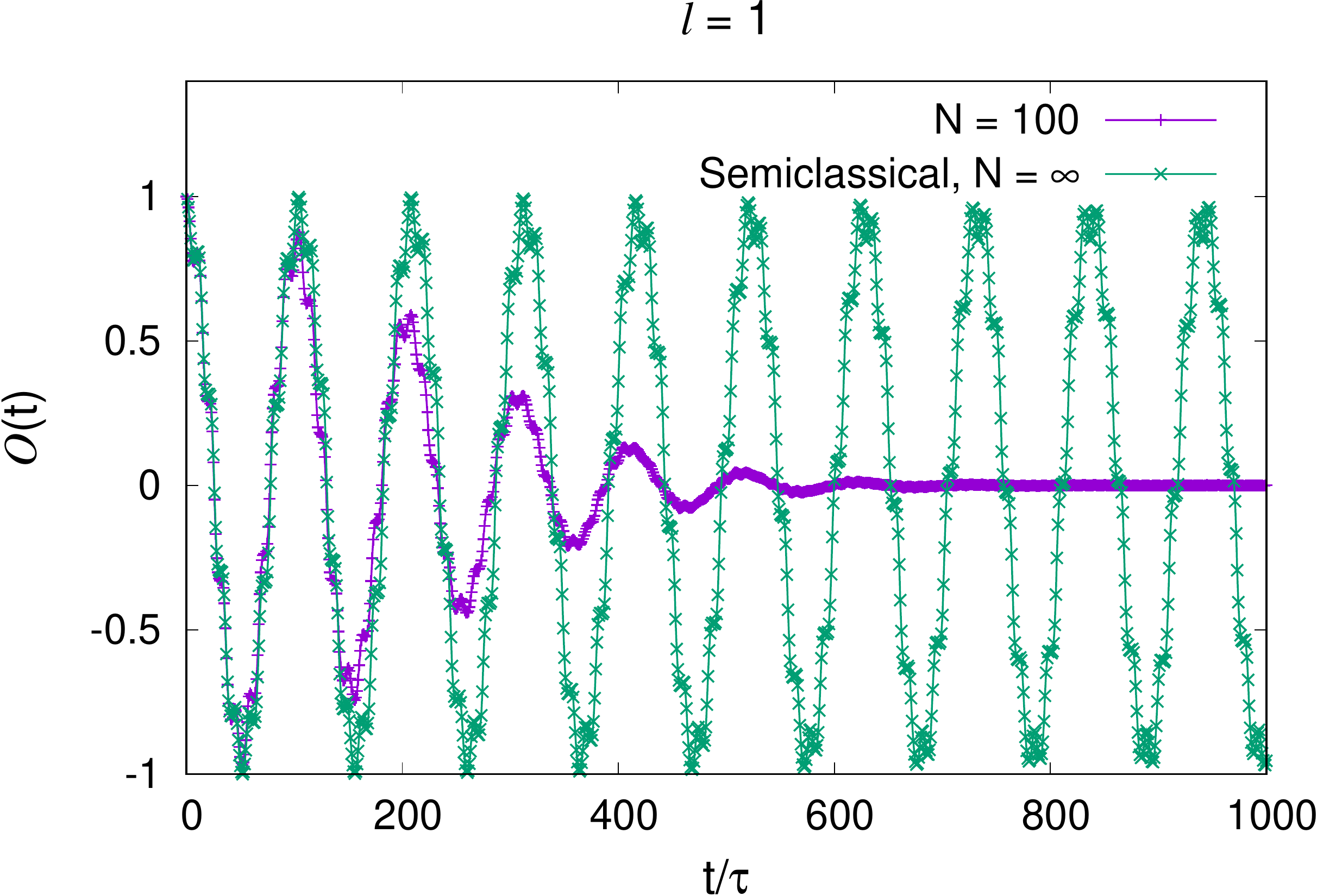}\put(20,57){(a)}\end{overpic}&
    \begin{overpic}[width=8cm]{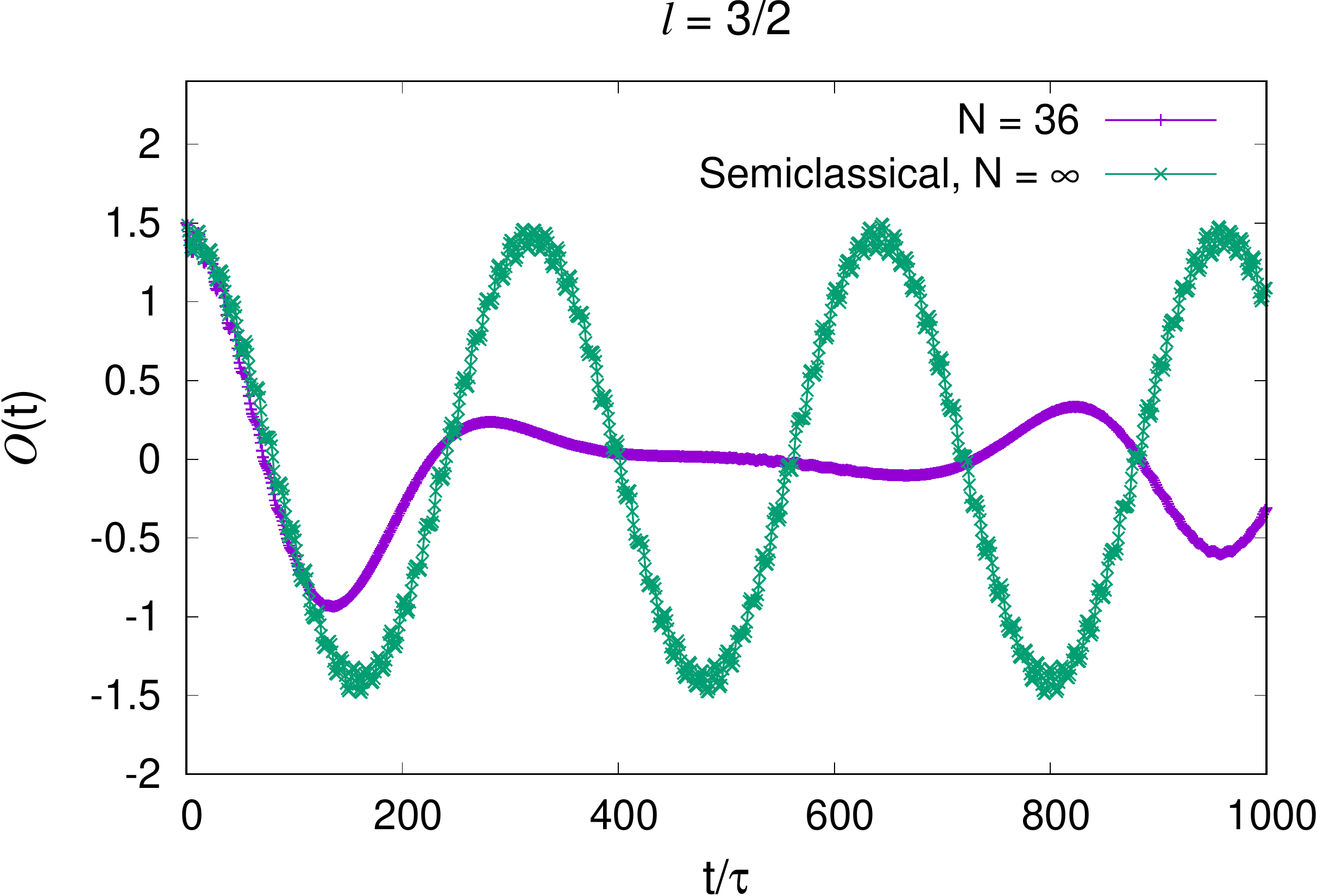}\put(20,57){(b)}\end{overpic}\\
    \begin{overpic}[width=8cm]{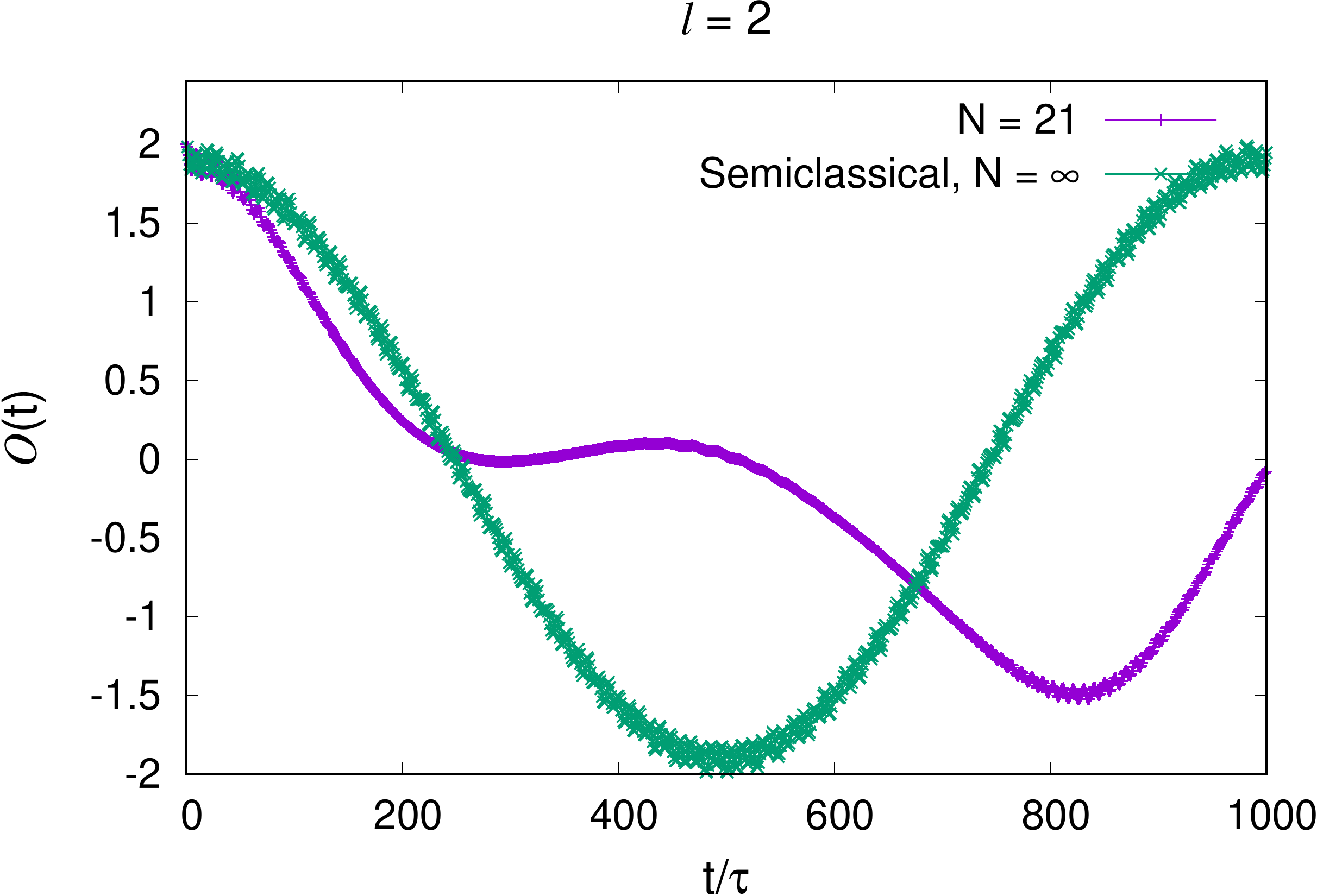}\put(20,57){(c)}\end{overpic}&
    \begin{overpic}[width=7cm]{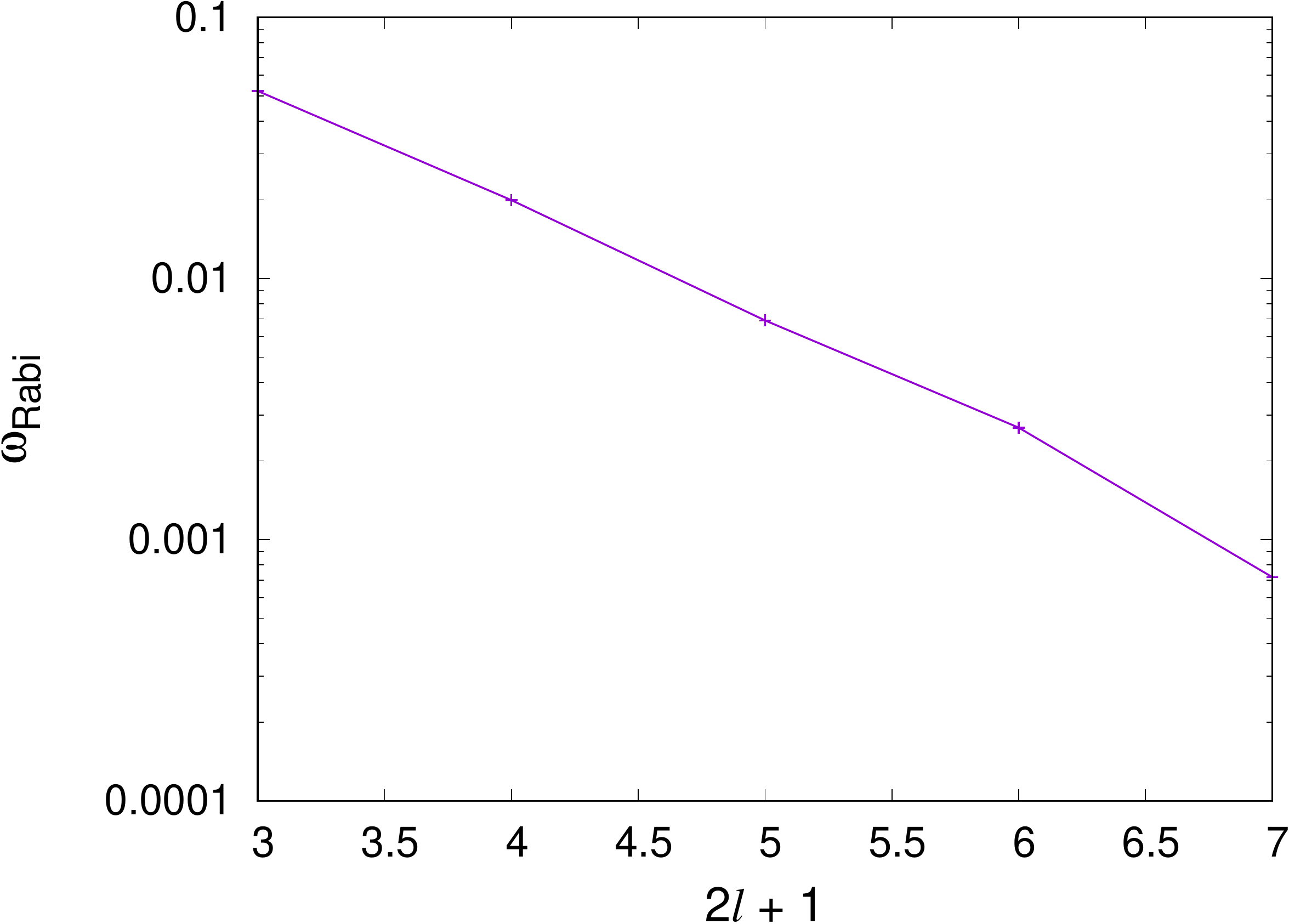}\put(80,60){(d)}\end{overpic}
  \end{tabular}
 \end{center}
 \caption{(Panels a-c) Stroboscopic evolution of $\mathcal{O}(t)$ versus $t/\tau$ for different values of $l$ and $N$ compared with the Gross-Pitaevskii $N\to\infty$ limit (same parameters as in Fig.~\ref{evo:fig}). Notice the Rabi oscillations in this limit which are washed out by quantum effects for $N$ finite. (Panel d) Rabi frequency of the $\mathcal{O}(t)$ oscillations versus $2l+1$. Numerical parameters: $h=0.1,\,\tau=0.6,\,J=1.0,\,K = 0.3,\,\phi=\pi$.}
    \label{semievo:fig}
\end{figure*}

We start from the Heisenberg equations for the operators $\opb{m,\,H}(t)$
\begin{widetext}
\begin{align}\label{Heisen_evo:eqn}
  &i\frac{\ud}{\ud t}\opb{m,\,H}(t)=-\frac{J}{l}m^2\opb{m,\,H}(t)-h\hat{\mathcal{A}}_H(t)+\delta_\tau(t)\left[\frac{\phi}{2}\hat{\mathcal{A}}_H(t)-\frac{K}{8Nl}\left\{\hat{\Sigma}_H(t),\hat{\mathcal{A}}_H(t)\right\}\right]\quad{\rm with}\nonumber\\
  &\hat{\mathcal{A}}\equiv\sqrt{l(l+1)-m(m+1)}\,\opb{m+1}+\sqrt{l(l+1)-m(m-1)}\,\opb{m-1}\quad{\rm and}\nonumber\\
  &\hat{\Sigma}\equiv\sum_{m=-l}^{l-1}\sqrt{l(l+1)-m(m+1)}\left(\opbdag{m}\,\opb{m+1} + \text{H.~c.}\right)\,.
\end{align}
\end{widetext}
We can write $\opb{m}=\sqrt{N}\hat{\beta}_m$. We see that 
\begin{equation}\label{bebeta:eqn}
  [\hat{\beta}_m,\hat{\beta}_m^\dagger]=\frac{1}{N}\,.
\end{equation}
 So, in the limit $N\to\infty$, these are classical variables and have vanishing correlations. Using this fact, evaluating the expectation over the initial state of Eq.~\eqref{Heisen_evo:eqn} (we define $\beta(t)\equiv\braket{\psi(0)|\hat{\beta}_{m,\,H}(t)|\psi(0)}$), and performing the limit $N\to\infty$, we get the Gross-Pitaevskii equations
\begin{widetext}
\begin{align}
  &i\frac{\ud}{\ud t}\beta_m(t)=-\frac{J}{l}m^2\beta_{m}(t)-h[\sqrt{l(l+1)-m(m+1)}\,\beta_{m+1}(t)+\sqrt{l(l+1)-m(m-1)}\,\beta_{m-1}(t)]\nonumber\\
  &+\delta_\tau(t)\left\{\frac{\phi}{2}-\frac{K}{2l}\left[\sum_{m'=-l}^{l-1}\sqrt{l(l+1)-m'(m'+1)}\Real\left(\beta_{m'}^*(t)\beta_{m'+1}(t)\right)\right]\right\}\nonumber\\
  &\cdot[\sqrt{l(l+1)-m(m+1)}\,\beta_{m+1}(t)+\sqrt{l(l+1)-m(m-1)}\,\beta_{m-1}(t)]\nonumber\\
\end{align}
\end{widetext}
with $\beta_m(t)\equiv 0$ for $m<-l$ or $m>l$. These equations are pretty simple to simulate numerically even for quite large values of $l$~\cite{notello} and we do it using 4th order Runge-Kutta~\cite{NumericalRecipes}. The initialization is
%
  $\beta_m(0) = \delta_{m\,l}$\,. 
%
The expectation of the operator $\hat{S}^z/N$ (see Eq.~\eqref{sezz:eqn}) is easily written as
\begin{equation}\label{sozzo:eqn}
  s^z(t)=\sum_{m=-l}^l m |\beta_m(t)|^2\,.
\end{equation}
We show some examples of stroboscopic Gross-Pitaevskii evolution compared with the finite $N$ cases in Fig.~\ref{semievo:fig} (a-c). For $N\to\infty$, we see very clear Rabi oscillations of $\mathcal{O}(t)$ with no decay. These oscillations come from the resonance between the state with $z$ angular momentum $l$ ($\beta_m=\delta_{l,\,m}$) and the one with $z$ angular momentum $-l$ ($\beta_m=\delta_{-l,\,m}$). 

At finite $N$ these states are $\ket{\psi_\uparrow}=\frac{1}{\sqrt{N!}}(\opbdag{l})^N\ket{0}$ and $\ket{\psi_\downarrow}=\frac{1}{\sqrt{N!}}(\opbdag{-l})^N\ket{0}$ and correspond to $z$ angular momentum $Nl$ and $-Nl$, respectively.
When $K,h\ll 1$, we expect that these states are connected in perturbation theory at order $\sim 2l+1$, so the frequency $\omega_{\rm Rabi}$ of the Rabi oscillations of $\mathcal{O}(t)$ should be of order~\cite{russomanno2017floquet}
\begin{equation}
 \omega_{\rm Rabi}\sim\left(\frac{\max(h,K)}{J}\right)^{2l+1}\hspace{-0.5cm}=\nep^{-(2l+1)\log\left(\frac{J}{\max(h,K)}\right)}\,.
\end{equation}
From our numerics we find exactly this exponential scaling [see Fig.~\ref{semievo:fig}(d)]. We evaluate $\omega_{\rm Rabi}$ frequency by performing the Fourier transform of the signal of $s^z(t)$, finding the frequency $\omega_{\rm peak}$ corresponding to the maximum of the power spectrum and then evaluating $\omega_{\rm Rabi}=\pi-\omega_{\rm peak}$. The vanishing of $\omega_{\rm Rabi}$ for $l\to\infty$ implies the existence of persisting period-doubling oscillations in this limit, which is equivalent to the classical case (see Sec.~\ref{models:sec}). In agreement with that, for the parameters of Fig.~\ref{semievo:fig}(d), the classical case Eq.~\eqref{eq:cl_Hamiltonian} shows period doubling, as one knows from Ref.~\cite{khasseh2019many}. 
We further remark that the Rabi oscillations for uncoupled spins ($K=0$) in this same model have been already observed in Ref.~\cite{russomanno2017floquet}. 

\begin{figure}
 \begin{center}
  \begin{tabular}{c}
    \begin{overpic}[width=8cm]{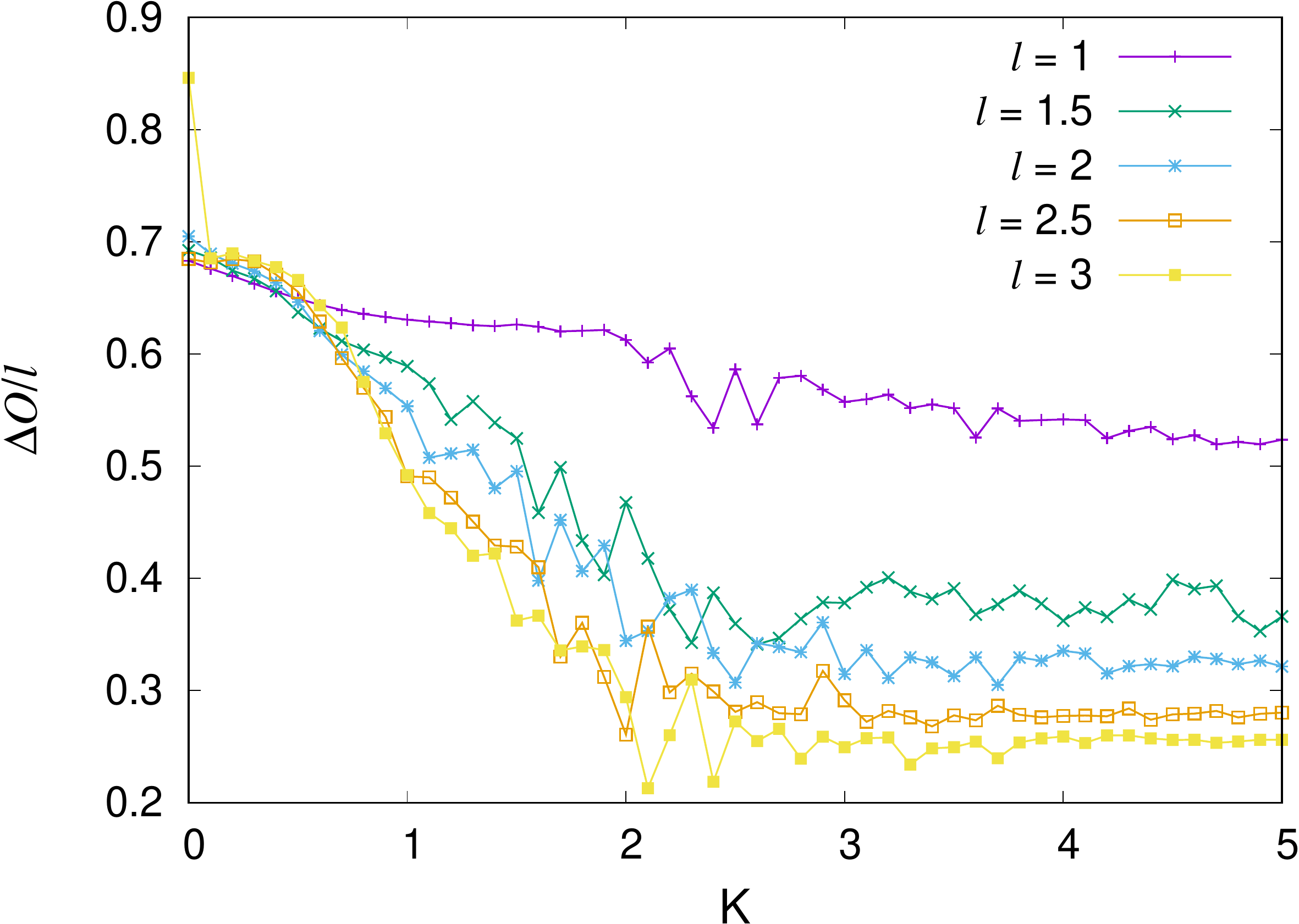}\put(20,57){(a)}\end{overpic}\\
    \begin{overpic}[width=8cm]{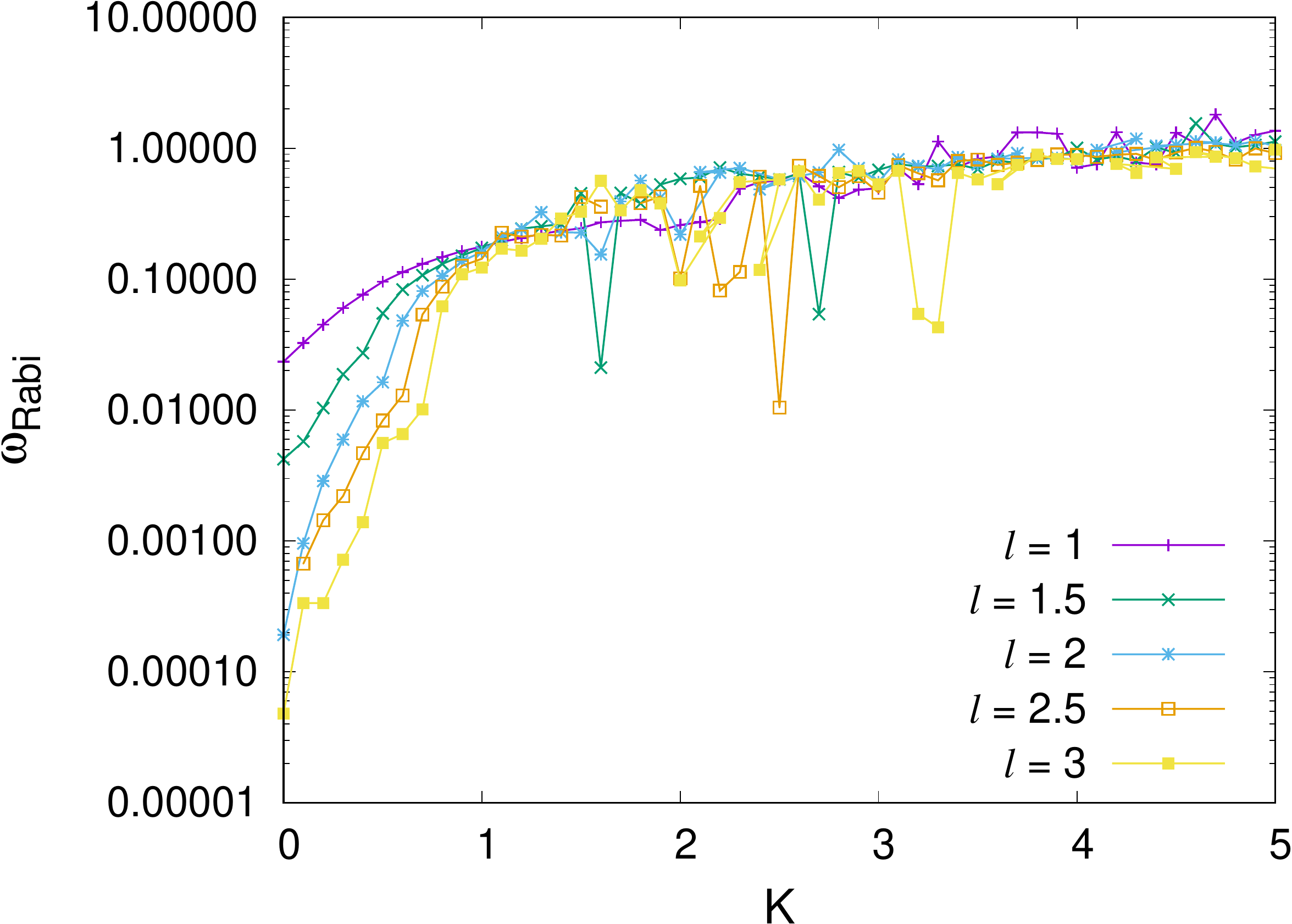}\put(30,61){(b)}\end{overpic}
  \end{tabular}
 \end{center}
 \caption{(Panel a) Amplitude of the Rabi oscillations versus $K$, (Panel b) $\omega_{\rm Rabi}$ versus $K$ for different values of $l$. Numerical parameters: $h=0.1,\,\tau=0.6,\,J=1.0,\,\phi=\pi$.}
    \label{crossing:fig}
\end{figure}

We consider also the amplitude of the Rabi oscillations $\Delta\mathcal{O}$. We define them as square deviation of $s^z(t)$ [Eq.~\eqref{sozzo:eqn}] over time. We call it $\Delta\mathcal{O}$ because it is also the mean square deviation of $\mathcal{O}(t)$, as it is easy to show. In order to make a comparison between different values of $l$ possible, we consider $\Delta\mathcal{O}/l$. We plot this quantity versus $l$ in Fig.~\ref{crossing:fig}(a). For every $l$, we see a crossing point between the curve for $l$ and the one for $l+0.5$. We see that the crossing moves towards the right for increasing $l$ and for $l=2$ the crossing is at $K^*\sim 0.7$. For $K<K^*$ the value of $\Delta\mathcal{O}/l$ increases with $l$, for $K>K^*$ it decreases. This suggests that there is a phase transition in the limit $l\to\infty$, as actually occurs~\cite{khasseh2019many}. Moreover, also the curves for $\omega_{\rm Rabi}$ versus $K$ show a crossing [Fig.~\ref{crossing:fig}(b)]. This crossing occurs for $K=1$ and there is no contradiction with the result for the amplitude because in that case $K^*$ increases with increasing $l$ and $K^*<1$. The crossing in the Rabi frequency is a strong evidence of a transition in the limit $l\to\infty$ between a period-doubling and a trivial phase, and corresponds to what is observed in the dynamics of Eq.~\eqref{eq:cl_Hamiltonian}.
%
\subsubsection{Largest Lyapunov exponent}\label{lyapunov:sec}
\begin{figure}
 \begin{center}
  \begin{tabular}{c}
    \includegraphics[width=8cm]{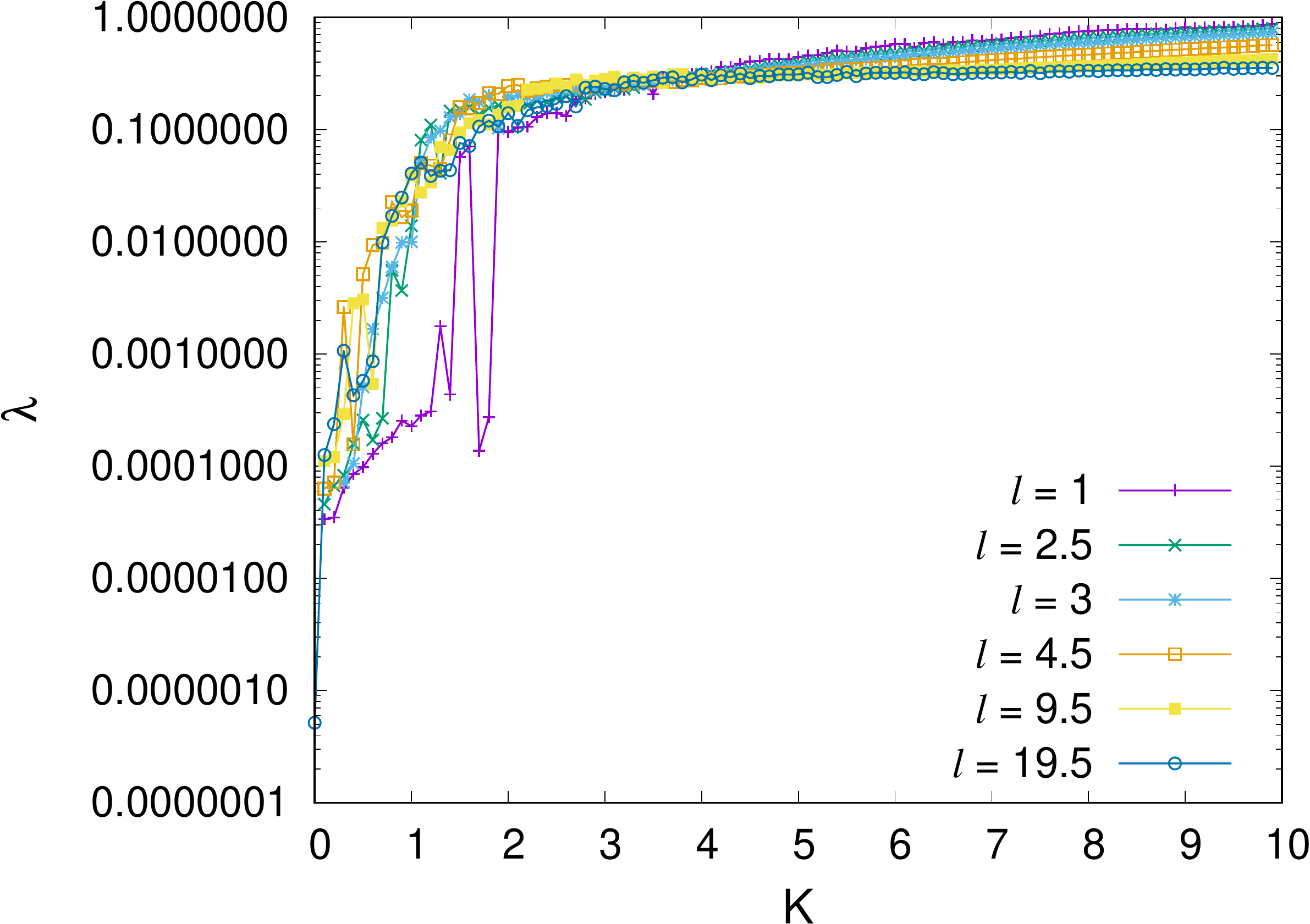}
  \end{tabular}
 \end{center}
 \caption{Largest Lyapunov exponent versus $K$ for different values of $l$. Numerical parameters: $h=0.1,\,\tau=0.6,\,J=1.0,\,K = 0.3,\,\phi=\pi,\,\mathcal{T}=2\cdot 10^{5}$. Initial distance between the two nearby initial conditions $d_0=10^{-10}$.}
    \label{lyapunov:fig}
\end{figure}

We evaluate here the largest Lyapunov exponent, which is a probe of exponential divergence of nearby trajectories and therefore a probe of chaotic dynamics~\cite{Ott:book}. The largest Lyapunov exponent is approximated as $\lambda(\mathcal{T})$, a stroboscopic average over $\mathcal{T}$ periods tending to $\lambda$ for $\mathcal{T}\to\infty$. We compute $\lambda(\mathcal{T})$ evaluating the rate of exponential increase in each period and averaging over periods. In practice, we consider two points in the phase space with distance $d_0$, we evolve over a period and consider the value of the distance $d_1$. Then we move the phase-space point of one of the trajectories along the segment joining the two so that we get again a distance $d_0\ll 1$, and evolve again for one period getting a distance $d_2$. Repeating $\mathcal{T}$ times, we get a sequence $\{d_n\}$ of distances~\cite{PhysRevA.14.2338} and we evaluate
$$
  \lambda(\mathcal{T})=\frac{1}{\mathcal{T}}\sum_{n=1}^{\mathcal{T}}\ln\left(\frac{d_n}{d_0}\right)\,.
$$
Taking $\mathcal{T}=2\cdot 10^{5}$ we already see convergence of $\lambda(\mathcal{T})$ and show the result in Fig.~\ref{lyapunov:fig}. What is remarkable is that this exponent is always positive, although it can get very small values ($<10^{-2}$) for $K<1$, marking thereby the existence of chaos. {This classical chaos is fully mirrored by the quantum chaos occurring for finite $N$ and appearing for any value of $K$ if $l$ is large enough (see Fig.~\ref{rorro:fig}). Only for $l=1$ and $K<2$ there is a lack of correspondence between the quantum behaviour (not quantum chaotic) and the classical nonvanishing Lyapunov exponent. Nevertheless, right at $K=2$ the Lyapunov exponent shows a discontinuity mirroring thereby the crossover in the quantum finite-$N$ behaviour.}

\begin{figure}
 \begin{center}
  \begin{tabular}{c}
    \includegraphics[width=8cm]{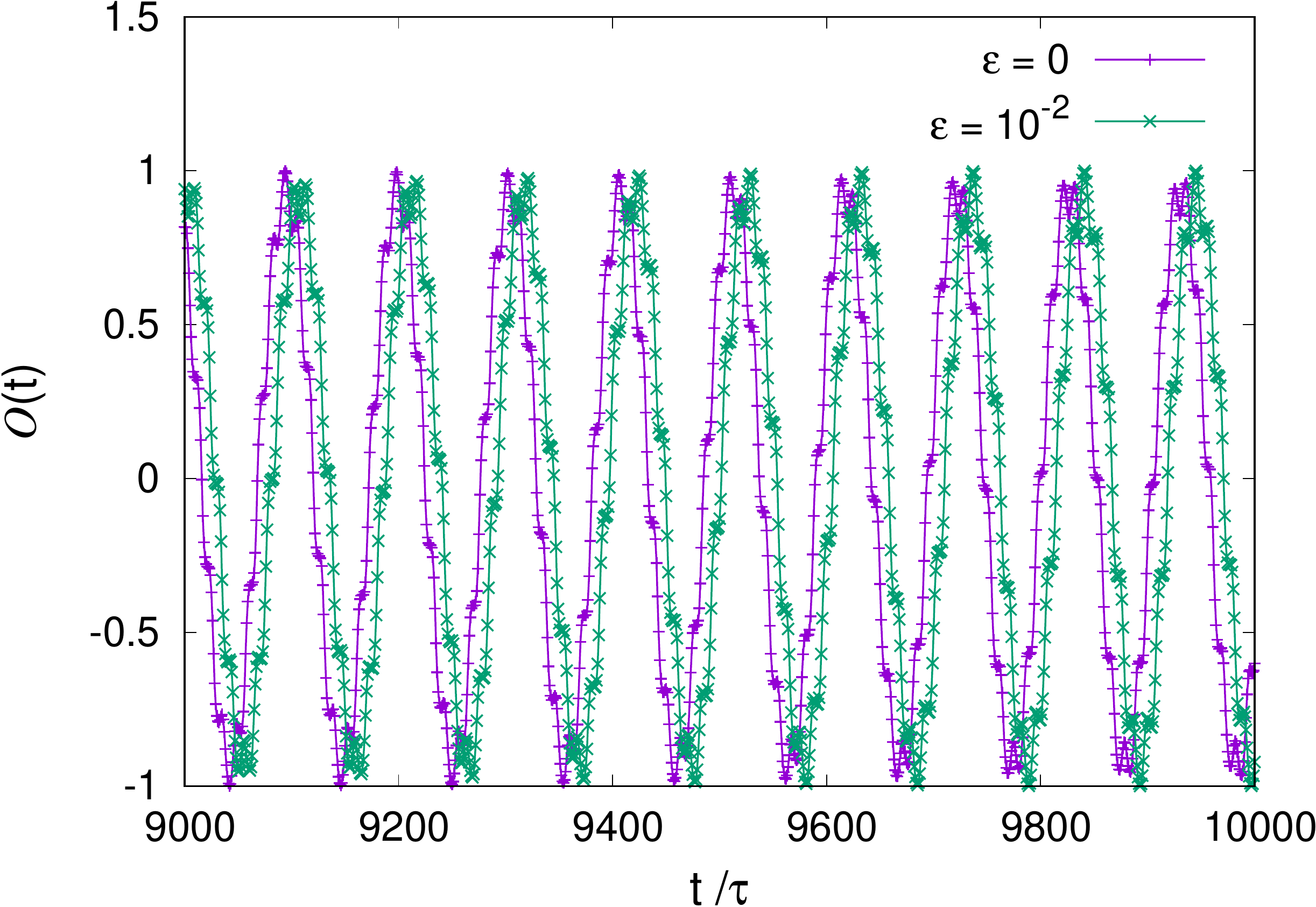}
  \end{tabular}
 \end{center}
 \caption{Same as Fig.~\ref{semievo:fig} (upper left panel) with two slightly different initializations. Numerical parameters: $h=0.1,\,\tau=0.6,\,J=1.0,\,K = 0.3,\,\phi=\pi$.}
    \label{difftra:fig}
\end{figure}

For small $K$, the system is chaotic but not ergodic. Indeed, it can support a regular behaviour as the one in Fig.~\ref{semievo:fig}. And we have checked that this behaviour is not due to an isolated regular trajectory: we see the same oscillations even if we take a slightly different initial state ($\beta_{m}(0)=\epsilon\delta_{m\,0}+\sqrt{1-\epsilon^2}\delta_{m\,1}$), see Fig.~\ref{difftra:fig}. Nevertheless, this is just a finite-time analysis and a chaotic behaviour might manifest at a time exponentially large in $1/K$~\cite{Nekhoroshev1971}.
  
The largest Lyapunov exponent plotted in Fig.~\ref{lyapunov:fig} allows to estimate the time scale over which the Gross-Pitaevskii description is valid for finite $N$. We see from Eq.~\eqref{bebeta:eqn} that for a finite-$N$ system the width of the quantum fluctuations of $\beta_m(t)$ is at best $\sim 1/\sqrt{N}$. Due to chaotic dynamics, this initial uncertainty increases exponentially in time with rate $\lambda$. The time the uncertainty reaches order 1 is
\begin{equation}
  t\sim\frac{1}{2\lambda}\log N\,.
\end{equation}
After this time, the dynamics is quantum. 
%
\section{Analysis of model-2
} \label{sec:sec}
We get this model by applying the substitution Eq.~\eqref{substo:eqn} into Eq.~\eqref{eq:cl_Hamiltonian} and then multiplying the resulting Hamiltonian by $2l$. The resulting Hamiltonian is given in Eq.~\eqref{eq:Hamiltonian_qauntum}.
Similarly to what we have done above, we define
$$
  \hat{S}^z\equiv\frac{1}{2}\sum_{j=1}^N\sum_{m=1}^{2l}\hat{\sigma}_{i,\,m}^z
$$
and, in order to understand if there is period doubling, we study the evolution of the period-doubling order parameter Eq.~\eqref{perdo:eqn}. 
%
%
Our initial state is given by
\begin{equation}\label{inuno:eqn}
  \ket{\psi(0)}=\ket{\up,\,\up,\,\ldots,\,\up}\,.
\end{equation}
We notice that for $K=0$ this model reduces to the kicked Lipkin model of Ref.~\cite{russomanno2017floquet}. This model showed period doubling for $h$ and $\phi-\pi$ small enough. In particular, $\mathcal{O}(t)$ showed Rabi oscillations with a frequency $\sim(h/J)^{2l}$. In the limit $l\to\infty$ (which in that context was the thermodynamic limit) the frequency of these oscillations tended to 0 and the period-doubling order parameter $\mathcal{O}(t)/l$ persisted to keep a finite value up to $t\to\infty$. Now we couple many of these models with each other by means of the coupling $K$. As we have seen in the discussion for the model-1, which is equivalent to this one, this coupling is not strong enough to stabilize the order parameter to a value different from 0 for any finite $l$. At most, if $K$ is small enough, the order parameter still shows Rabi oscillations with a renormalized frequency.

 We study here the model-2 by means of the DTWA, an approximation which has proved to work fine in a long-range context~\cite{pappalardi2018scrambling,schachenmayer2015many,khasseh2020discrete}. We see that the DTWA is unable to reproduce the Rabi oscillations, but correctly gets the fact that there is no period doubling for finite $l$ in the limit of large $N$. We get period doubling, in agreement with the exact dynamics, only in the classical $l\to\infty$ limit. We briefly outline the DTWA approach in the next subsection. 
\subsection{Discrete truncated Wigner approximation in a nutshell}\label{dtwa:sec}
This is an approximation method especially convenient for long-range interacting spin models. All the details can be found in~\cite{pappalardi2018scrambling,schachenmayer2015many,reyhaneh}. Here we just outline the application to our case. We start by expanding the expectation of a generic operator $\hat{\mathcal{B}}$ in a basis of operators in the form
\begin{equation} \label{cacabo:eqn}
  \mean{\mathcal{B}}_t=\sum_{\bbeta}w_{\bbeta}\,\mathcal{B}_{\bbeta}(t)
\end{equation}
where $w_{\bbeta}\equiv\frac{1}{2}\Tr\left[\hat{A}_{\bbeta}\hat{\rho}\right]$ is the Wigner function, $\mathcal{B}_{\bbeta}^w(t)=\Tr\left[\hat{A}_{\bbeta}\hat{\mathcal{B}}(t)\right]$ 
are the Weyl symbols and $\hat{\mathcal{B}}(t)\equiv\hat{U}^\dagger(t,0)\hat{\mathcal{O}}\hat{U}(t,0)$ with $\hat{U}(t,0)$ the time-evolution operator form 0 to $t$ of the Hamiltonian Eq.~\eqref{eq:Hamiltonian_qauntum}. We can take a basis of operators factorized over the sites
\begin{equation}
  \hat{\mathcal{A}_{\bbeta}}=\bigotimes_{j,\,m}\mathcal{A}_{\beta_{j,\,m}}
\end{equation}
where we can take over each site~\cite{Wootters}
\begin{equation} \label{ini:eqn}
  \hat{A}_{\beta}=\frac{\boldsymbol{1}+\textbf{s}_{\beta}\cdot\hat{\boldsymbol{\sigma}} }{2}
\end{equation}
where $\boldsymbol{s}_\beta$ can take the values $\left(\begin{array}{ccc}1&1&1\end{array}\right)$, $\left(\begin{array}{ccc}-1&1&-
1\end{array}\right)$, $\left(\begin{array}{ccc}1&-1&-1\end{array}\right)$ and $\left(\begin{array}{ccc}-1&-1&1\end{array}\right)$ and 
$\hat{\boldsymbol{\sigma}}=\left(\begin{array}{ccc}\hat{\sigma}^x&\hat{\sigma}^y&\hat{\sigma}^z\end{array}\right)$. The approximation amounts to take the evolution of $\hat{\mathcal{A}_{\bbeta}}$ as factorized
\begin{equation}
  \hat{U}(t,0)\hat{\mathcal{A}_{\bbeta}}\hat{U}^\dagger(t,0)=\bigotimes_{j,\,m}\mathcal{A}_{\beta_{j,\,m}}(t)
\end{equation}
with
\begin{equation}
  \hat{A}_{\beta_{j,\,m}}(t)=\frac{\boldsymbol{1}+\sum_{\mu=x,\,y,\,z}{s}_{j,\,m,\,\beta_{j,\,m}}^\mu(t)\hat{\sigma}_{j,\,m}^\mu}{2}\,.
\end{equation}
The ${s}_{j,\,m,\,\beta_{j,\,m}}^\alpha(t)$ have as initial values the ones given in Eq.~\eqref{ini:eqn}, for the corresponding $\beta_{j,\,m}$, and obey the evolution equations
\begin{align}
  &\dot{s}_{j,\,m,\,\beta_{j,\,m}}^{\mu}(t)
     =-\{s_{j,\,m,\,\beta_{j,\,m}}^{\mu}(t),{\cal H}^{(2)}\}\nonumber\\
       &=2\sum_{\nu,\,\rho=x,\,y,\,z}\epsilon^{\mu\nu\rho}s_{j,\,m,\,\beta_{j,\,m}}^{\rho}(t)\frac{\partial{\cal H}^{(2)}}{\partial s_{j,\,m,\,\beta_{j,\,m}}^{\nu}}\,.
\end{align}
where $\epsilon^{\mu\nu\rho}$ is the usual Ricci tensor, the elementary Poisson brackets are $\left\{s_{j,\,m,\,\beta_{j,\,m}}^\mu,\,s_{i,\,m',\,\beta_{i,\,m'}}^\nu\right\}=\epsilon^{\mu\,\nu\,\rho}\delta_{i\,j}\delta_{m\,m'}s_{j,\,m,\,\beta_{j,\,m}}^\rho$ and we have defined
\begin{equation}\label{eq:Hamiltonian_qauntum}
\begin{split}
&\mathcal{H}^{(2)}=-\sum_{i=1}^{N}\Big[\frac{J}{4l}\hspace{-0.2cm}\sum_{m,m'=1}^{2l}\hspace{-0.3cm}s_{i,\,m,\beta_{i,\,m}}^zs_{i,\,m,\beta_{j,\,m'}}^z+h\sum_{m=1}^{2l}\hspace{-0.1cm}s_{i,\,m,\beta_{i,\,m}}^x\Big] \\
&+\delta_{\tau}(t)\Big[\frac{\phi}{2}\sum_{m=1}^{2l}\hspace{-0.1cm}s_{i,\,m,\beta_{i,\,m}}^x\hspace{-0.2cm}-\frac{K}{16N\,l}\sum_{i,j\neq i}\,\sum_{m,m'=1}^{2l}\hspace{-0.3cm}s_{i,\,m,\beta_{i,\,m}}^xs_{j,\,m,\beta_{j,\,m}}^x\Big]\,.
\end{split}  
\end{equation}
In our case we can implement a Monte Carlo sampling procedure to approximate the sum of $4^N$ terms in Eq~\eqref{cacabo:eqn} in a numerically feasible way. Being the initial state Eq.~\eqref{inuno:eqn} given by the density matrix
\begin{equation}
  \hat{\rho}(0)=\bigotimes_j\frac{1}{2}\left(\mathcal{A}_{(-1\,-1\,1)}+\mathcal{A}_{(1\,1\,1)}\right)\,,
\end{equation}
in Eq.~\eqref{cacabo:eqn} we have that $w_{\bbeta}=1/2^N$ for all the products of operators in Eq.~\eqref{ini:eqn} containing only $\beta_{j,\,m}=(1\,-1\,-1)$ and $\beta_{j,\,m}=(1\,1\,1)$. So, one can approximate the expectation of any operator with a Monte Carlo sampling of the uniform distribution $w_{\bbeta}$, with the desired accuracy. More specifically, we focus on the expectation
\begin{equation}
  \braket{\psi(t)|\hat{S}^z|\psi(t)}=\sum_{\bbeta}w_{\bbeta}\sum_{j=1}^N\sum_{m=1}^{2l}s_{i,\,m,\,\beta_{i,\,m}}^z
\end{equation}
and we evaluate it as the average over $n_r$ random 
initializations where each $s_{j,\,\beta_j}$ is initialized with probability $1/2$ in the condition $\left(\begin{array}{ccc}1&1&1\end{array}\right)$ and probability $1/2$ in 
the condition $\left(\begin{array}{ccc}-1&-1&1\end{array}\right)$. Remarkably, the error bars do not scale with the system size, so this method is feasible also in the case of 
large systems~\cite{reyhaneh}. The errorbars are evaluated as $1/\sqrt{n_r}$ times the mean square deviation over randomness. In our analysis we have found that already for $n_r=800$ we have a satisfying convergence (see Appendix~\ref{app2:sec}). We are going to apply the DTWA method in the next subsection to study the period-doubling dynamics of the model-2.
\subsection{Results} \label{reso:sec}
\begin{figure}
\begin{overpic}[width=88mm]{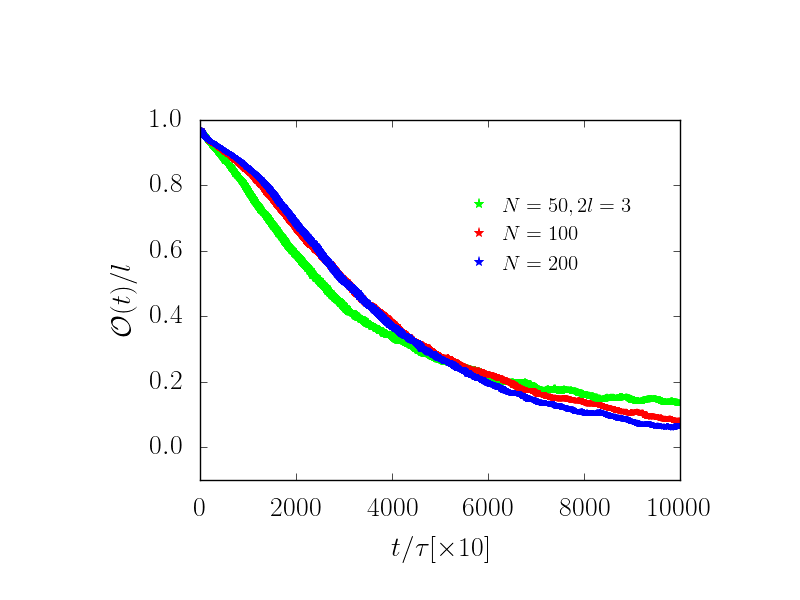}\put(-1,69){(a)}\end{overpic}
\begin{overpic}[width=88mm]{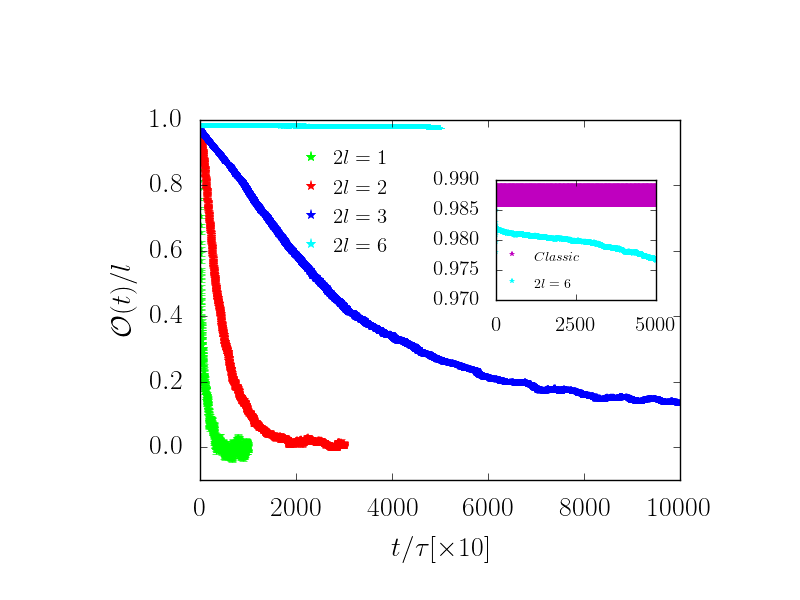}\put(-1,69){(b)}\end{overpic}
\caption{Dynamics of order parameter for (a) $2l=3$ and (b) different $2l$ and $N=50$. 
Numerical parameters: $h=0.1, \phi=\pi$, $K=0.3, \tau=0.6, $ and $n_r=800$.}
\label{fig2}
\end{figure}

Consistently with the results found in the case of the model-1 (Sec.~\ref{fir:sec}) we find here no period doubling. Indeed the period doubling order parameter $O(t)$ decays to 0 in a finite time, independent of the system size $N$ [see an example for $l=3/2$ in Fig.~\ref{fig2}(a)], for a set of parameters where the classical model Eq.~\eqref{eq:cl_Hamiltonian} shows period doubling. As in the model-1, the limit $l\to\infty$ corresponds to the classical case Eq.~\eqref{eq:cl_Hamiltonian}. We show this fact in Fig.~\ref{fig2}(b) where we fix $N$ and show the stroboscopic evolution of $O(t)$ versus $t/\tau$ for different values of $l$. We qualitatively see that $O(t)$ decay over a longer time as $l$ increases. We plot for comparison also the stroboscopic evolution of $(-1)^{t/\tau}\frac{1}{N}\sum_j m_j^z(t)$ in the classical case Eq.~\eqref{eq:cl_Hamiltonian}. This quantity persists for an infinite time and $O(t)$ tends to this curve when $l\to\infty$. We notice that already for $l=3$ the quantum dynamics is very near to the classical one, at least until $t/\tau=4\cdot 10^3$.

\begin{figure}
	\begin{tabular}{c}
		\begin{overpic}[width=60mm]{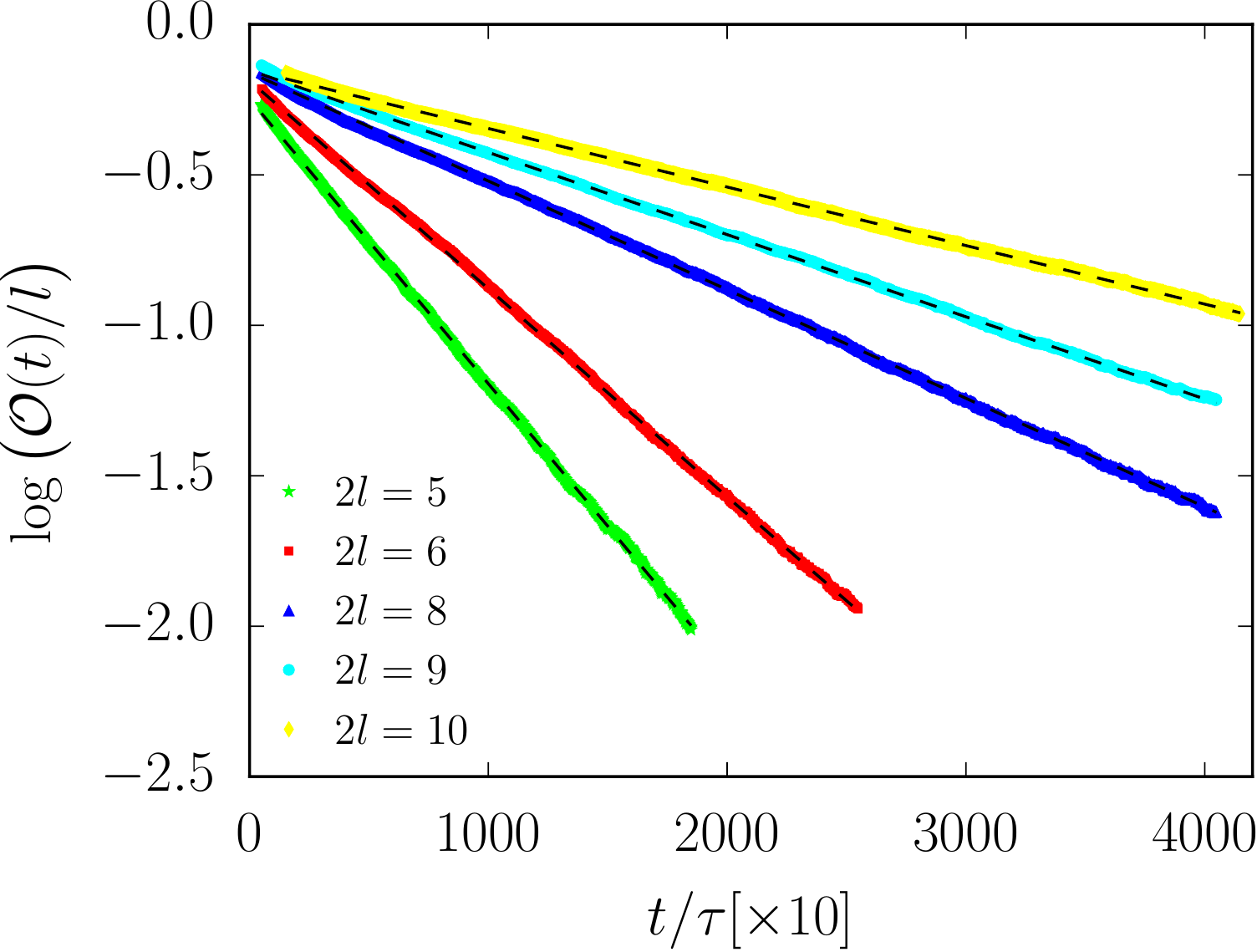}\put(-1,69){(a)}\end{overpic}\\
		\begin{overpic}[width=60mm]{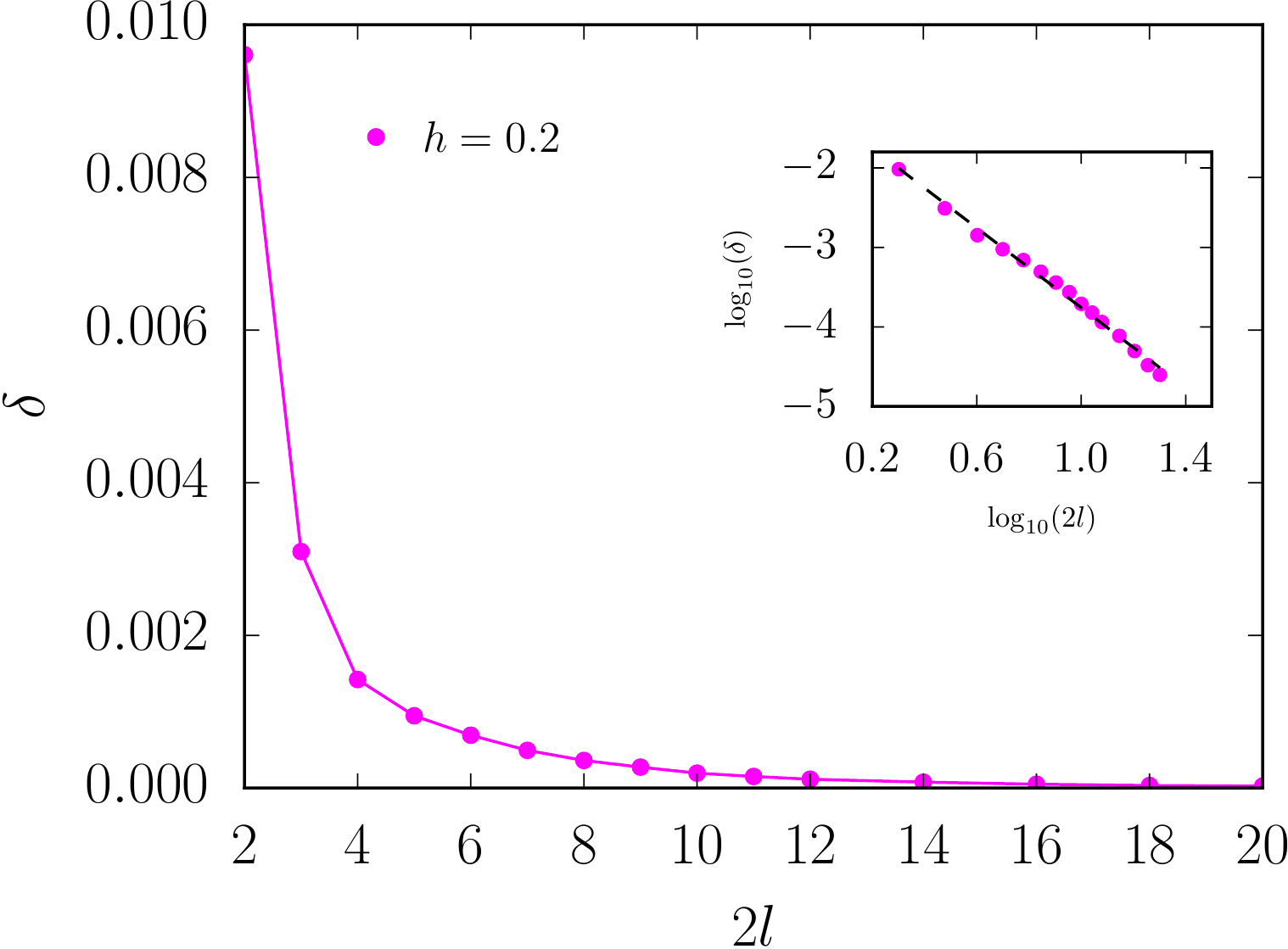}\put(-1,69){(b)}\end{overpic}
	\end{tabular}
	\caption{(a) Dynamics of the order parameter as a function of $t/\tau$. (b) $\delta$ exponent as a function of domain length $2l$. Numerical parameters: $N=50, h=0.2, \phi=\pi$, $K=0.3, \tau=0.6, $ and $n_r=800$.}
	\label{fig5}
\end{figure}

Let us move to study the decay of the period-doubling order parameter in a more quantitative way. First of all, we plot $\mathcal{O}(t)/l$ versus $t/\tau$ with a logarithmic scale along the vertical axis [see Fig.~\ref{fig5}(a)] and we see that $\mathcal{O}(t)/l$ decays exponentially in time. We can find the rate of this decay by fitting the curve of $\log O(t)$ versus $t$ with a straight line of the form $\log[ \mathcal{O}(t)/l]=A-\delta \,t$. We plot $\delta$ versus $l$ in Fig.~\ref{fig5}(b). We see that $\delta$ decays with $l$ as a power law, $\delta\sim 1/l^{\gamma}$. Fitting the bilogarithmic plot with a straight line we find the decay exponent to be $\gamma\simeq 2.51$. So, extrapolating, we find that $\delta\to 0$ when $l\to\infty$ and so in this limit the classical model and the period doubling are recovered. In Appendix~\ref{app2:sec} we discuss another method to estimate the decay time of $O(t)$ which gives similar results.

We remark that the model-2 is a different representation of the model-1 for the chosen initialization, as we have discussed in Sec.~\ref{models:sec}. DTWA therefore gives results which are not quantitatively correct (it does not catch any Rabi oscillation) and become so only in the limit $l\to\infty$. Nevertheless, when $l$ is finite, this approximation correctly gets the absence of persistent period doubling in the limit of large $N$.
\section{Conclusion}
In conclusion we add quantum fluctuations to a classical and Hamiltonian model of interacting classical angular momenta showing synchronized period doubling in the thermodynamic limit. We consider the case of all-to-all interactions where the long range correlations are stronger and the synchronized period doubling is most robust in the classical case. We study the robustness of synchronized period doubling adding quantum fluctuations in two different ways, realizing two different quantum models. In both the quantum models we find that the synchronized period doubling is fragile to quantum fluctuations and disappears. We perform our analysis by means of the so-called period doubling order parameter. When the system shows period doubling in the thermodynamic limit, the first zero of this order parameter occurs at at a time $t^*$ scaling to infinity for increasing system size. In both the quantum models $t^*$ does not increase with the system size, and so there is no period doubling, whenever the quantum fluctuations are significant.

We construct the quantum model-1 by replacing the classical angular momenta with quantum spins of size $l$. For any finite $l$ there are quantum fluctuations and we show that the model becomes classical in the limit $l\to\infty$. We restrict to the subspace even under all the permutation symmetries of the Hamiltonian, performing a mapping over a bosonic model. Due to the moderate Hilbert subspace dimension, we perform exact diagonalization for quite large system sizes and do the finite-size scaling of $t^*$. For all the accessible values of $l$, we find no scaling, and so there is no period doubling. 

This result is confirmed in the limit of infinite system size ($N\to\infty$), where the bosonic model is described by a system of Gross-Pitaevskii equations. In this limit, the period-doubling order parameter performs Rabi oscillations related to the existence of resonant states in the spin model. For increasing $l$, these states are connected at higher orders in perturbation theory and consistently the frequency of the Rabi oscillations exponentially decreases in $l$. In particular, for $l\to\infty$ the Rabi frequency goes to 0, so $t^*$ tends to infinity, and the classical period doubling is restored, consistently with the model becoming classical in this limit. Studying the dependence of the amplitude and the frequency of the Rabi oscillations on the parameter $K$,  we find that the curves for different $l$ cross at a point around $K\sim 1$. This point corresponds to the transition from synchronized to unsynchronized behaviour in the classical $l\to\infty$ limit.

For any finite $N$ and the accessible values of $l$, we observe quantum chaos in this model, as shown by the average level spacing ratio being Wigner-Dyson. This is true for $K\gtrsim 2$ for $l=1$ and for any value of $K$ for $l=2$. Analogously, in the $N\to\infty$ limit, the Gross-Pitaevskii equations show a positive largest Lyapunov exponent $\lambda$ (with a discontinuity at $K=2$ for $l=1$) and then the dynamics are chaotic. The Lyapunov exponent spans many orders of magnitude as $K$ increases. In particular, for the values $K<1$ corresponding to a synchronized period doubling in the limit $l\to\infty$, we see $\lambda\leq 10^{-2}$ for all the considered values of $l$. In this case we have a chaotic but not ergodic dynamics, as we verify by checking the existence of the Rabi oscillations of the period doubling order parameter also for a different initial condition. For any finite size $N$, we show that the dynamics is correctly described by the Gross-Pitaevskii equations up to a time scaling with $\log N$.

Then we move to introducing the model-2. We substitute the classical angular momenta with sums of $2l$ Pauli matrices. We argue that also this model reduces to the classical one in the limit $l\to\infty$. We find convenient to study this model by an approximation method called DTWA, which is known to give good results for long-range interacting models, and describe it in some detail. We focus on a set of parameters where the classical model shows period doubling  and we use DTWA to study the evolution of the period-doubling order parameter. We find that it decays to zero as an exponential and the time scale of this decay does not scale with the system size, marking the absence of period doubling. Nevertheless, the time scale of the exponential decay scales as a power law with $l$. So, in the limit of $l\to\infty$ there is period doubling, consistently with the model being classical in this limit. This model is equivalent to model-1 for the chosen initial conditions, so we see that DTWA provides quantitatively correct results only for $l\to\infty$. For $l$ finite it is not correct (it does not provide Rabi oscillations) but correctly predicts the absence of persistent period doubling in the limit of large $N$.

Therefore, we find that the period doubling in this model is fragile to the smallest quantum fluctuations. Prospects of future research include the exploration of Hamiltonian synchronization in different models, with stronger long-range correlations, and the study of its stability under quantum and thermal fluctuations.
\acknowledgements{Part of the computational resources were provided by CINECA on the Marconi-A3 partition through the HPC agreement between CINECA and the ICTP. We acknowledge T.~Mendes who provided us the access to these resources.}
\appendix
%
\section{Limit $l\to\infty$ as a classical limit}\label{app:lim}
Due to the equivalence of the two models for the chosen initial conditions, we discuss only the model-1. Conclusions apply also to the second one, not in general but for the chosen initial conditions. The analysis strictly resembles the one leading to the Gross-Pitaevskii for the bosonic model in the $N\to\infty$, as we have discussed in Sec.~\ref{limit:sec}. We rescale the spin variables as $\hat{\mathcal{S}}_j^\alpha=\frac{1}{l}\hat{s}_j^\alpha$. Their commutator is $[\hat{\mathcal{S}}_j^\alpha,\,\hat{\mathcal{S}}_i^\beta]=\frac{i}{l}\epsilon^{\alpha\beta\gamma}\hat{\mathcal{S}}_j^\gamma\delta_{i\,j}$, so these variables are classical in the limit $l\to\infty$. One can write the Heisenberg equations for the $\hat{\mathcal{S}}_j^\alpha$ variables
\begin{equation}
  \frac{\ud}{\ud t}\hat{\mathcal{S}}_j^\alpha(t)=[\hat{\mathcal{S}}_j^\alpha(t),\,\hat{H}^{(1)}(t)]\,,
\end{equation}
then evaluate the expectation over the initial state and, performing the limit $l\to\infty$, neglect any quantum correlation due to the vanishing commutator. Performing this calculation one gets $l\to\infty$ evolution equations for the expectations of $\hat{\mathcal{S}}_j^\alpha$ which exactly coincide with the classical evolution equations for $m_j^\alpha(t)$ obtained with Eq.~\eqref{eq:cl_Hamiltonian} using the classical Poisson brackets.

There is another method, nearer to the analysis of~\cite{Sciolla_2}, which we are going to sketch. Take for instance the operator $\hat{s}_j^x$ and apply it to the state $\ket{l,\,m}_j$, the eigenstate with eigenvalue $l(l+1)$ of $\hat{\boldsymbol{s}}_j^2$ and eigenvalue $m$ of $\hat{s}_j^z$. One gets~\cite{Picasso:book}
\begin{align}
  \hat{s}_j^x\ket{l,\,m}_j &= \frac{1}{2}\Big(\sqrt{l(l+1)-m(m+1)}\ket{l,\,m+1}_j\nonumber\\
   &+\sqrt{l(l+1)-m(m-1)}\ket{l,\,m-1}_j\Big)\,.
\end{align}
Defining $q_j=m/l$ and using the translation operator of shift $a$ along $q_j$, $\exp(a\frac{\ud}{\ud q_j})$, one finds for $l\gg 1$
\begin{equation}
  \hat{\mathcal{S}}_j^x=\frac{1}{l}\hat{s}_j^x\ket{l,\,m}_j \simeq\sqrt{1-q_j^2}\cos\left(\frac{1}{l}\frac{\ud}{\ud q_j}\right)\ket{l,\,m}_j\,.
\end{equation}
Defining $\hat{p}_j=-\frac{i}{l}\frac{\ud}{\ud q_j}$, we see that we have written this object in terms of two canonical variables, $\hat{q}_j$ and $\hat{p}_j$ whose commutator is $[\hat{q}_j,\,\hat{p}_j]=i/l$. In the limit $l\to\infty$, therefore, they are classical canonical variables obeying the canonical Poisson bracket $\{q_j,\,p_j\}=-1$. The classical $m_j^\alpha$ can then be obtained as $m_j^x=b\sqrt{1-q_j^2}\cos p_j$, with $b>0$ arbitrary giving the size of the classical spin. In the same way one gets~\cite{Sciolla_2} $m_j^z=b q_j^2$ and $m_j^y=b\sqrt{1-q_j^2}\sin p_j$. Appropriately fixing $b=1/2$ one gets the classical Hamiltonian Eq.~\eqref{eq:cl_Hamiltonian}. The canonical Poisson brackets of $q_j$ and $p_j$ give rise to the angular momentum Poisson brackets for the $m_j^\alpha$ which are stated immediately below Eq.~\eqref{eq:cl_Hamiltonian}. In this way one gets back the classical dynamics for $l\to\infty$.
\section{Mapping onto the bosonic model} \label{boso_map:sec}
We can now discuss the bosonic mapping of the notes. This mapping was introduced in~\cite{federica,PhysRevB.103.224301} for similar infinite-range models. Let us consider a system of $N$ sites, and let us take the local spins with value $l=1$ for clarity (the generic case is exactly identical). With $l=1$, we can have $m=-1,\,0,\,1$. Because the system is fully symmetric under permutations, we can restrict to the states even under all the possible $N!$ permutations. If we call $\hat{P}$ the sum of all the permutation operators, we can take as basis of our Hilbert space the states
\begin{align} \label{defio:eqn}
  &\ket{n_{-1}\,n_0\,n_1}\nonumber\\
  &\equiv\frac{1}{\sqrt{N!\,(n_{-1}!n_0!n_1!)}}\hat{P}\ket{\underbrace{(-1\ldots-1)}_{n_{-1}}\underbrace{(0\ldots0)}_{n_{0}}\underbrace{(1\ldots1)}_{n_{1}}}\,.\nonumber\\
\end{align}
There are $N$ sites, so $n_{-1}+n_0+n_1=N$. The factor in front is for normalization, and the $\sqrt{N!}$ is there because there are $N!$ possible permutations. The factors $\sqrt{n_m!}$ at the denominator are there for the following reason. Consider for instance $m=1$. Fixing everything else, there are $n_1!$ ways of rearrange the sites with $m=1$ and this increases the norm by a factor $n_1!$. One divides by $\sqrt{n_1!}$ and the norm is again 1.

Consider for instance the application of the operator $\hat{S}^+=\sum_j\hat{s}_j^+$, where $\hat{s}_j^+=\hat{s}_j^x+i\hat{s}_j^y$. One finds
\begin{align}
  &\hat{S}^+\ket{n_{-1}\,n_0\,n_1}\nonumber\\
  &=\frac{1}{\sqrt{N!\,(n_{-1}!n_0!n_1!)}}\hat{P}\sum_j\hat{s}_j^+\ket{\underbrace{(-1\ldots-1)}_{n_{-1}}\underbrace{(0\ldots0)}_{n_{0}}\underbrace{(1\ldots1)}_{n_{1}}}\,.\nonumber\\
\end{align}
If for instance $\hat{s}_j^+$ acts over a site with $m=-1$, one gets a factor $\sqrt{1(1+1)}=\sqrt{2}$~\cite{Picasso:book}. Moreover, there are $n_{-1}$ of these sites and they are equivalent due to the permutation operator. This gives rise to a factor $n_{-1}$ in front. Moreover, in this way one decreases $n_{-1}$ by 1 and increases $n_{0}$ by 1. One has a similar situation for $m=0$ and $m=1$, so
\begin{widetext}
\begin{align}
  \hat{S}^+\ket{n_{-1}\,n_0\,n_1}&=\frac{1}{\sqrt{N!\,(n_{-1}!n_0!n_1!)}}n_{-1}\sqrt{2}\,\hat{P}\ket{\underbrace{(-1\ldots-1)}_{n_{-1}-1}\underbrace{(0\ldots0)}_{n_{0}+1}\underbrace{(1\ldots1)}_{n_{1}}}\nonumber\\
 &+\frac{1}{\sqrt{N!\,(n_{-1}!n_0!n_1!)}}n_{0}\sqrt{2}\,\hat{P}\ket{\underbrace{(-1\ldots-1)}_{n_{-1}}\underbrace{(0\ldots0)}_{n_{0}-1}\underbrace{(1\ldots1)}_{n_{1}+1}}\,.
\end{align}
\end{widetext}
Doing some algebra, one can write
\begin{widetext}
\begin{align}
  \hat{S}^+\ket{n_{-1}\,n_0\,n_1}&=\frac{\sqrt{n_{-1}(n_0+1)}\sqrt{2}}{\sqrt{N!\,((n_{-1}-1)!(n_0+1)!n_1!)}}\,\hat{P}\ket{\underbrace{(-1\ldots-1)}_{n_{-1}-1}\underbrace{(0\ldots0)}_{n_{0}+1}\underbrace{(1\ldots1)}_{n_{1}}}\nonumber\\
 &+\frac{\sqrt{n_{0}(n_1+1)}\sqrt{2}}{\sqrt{N!\,(n_{-1}!(n_0-1)!(n_1+1)!)}}\,\hat{P}\ket{\underbrace{(-1\ldots-1)}_{n_{-1}}\underbrace{(0\ldots0)}_{n_{0}-1}\underbrace{(1\ldots1)}_{n_{1}+1}}\,.
\end{align}
\end{widetext}
Using the definition~\eqref{defio:eqn} one can write
\begin{align}
  &\hat{S}^+\ket{n_{-1}\,n_0\,n_1}\nonumber\\
  &=\sqrt{2}\sqrt{n_{-1}(n_0+1)}\ket{n_{-1}-1,\,n_0+1,\,n_1}\nonumber\\
  &+\sqrt{2}\sqrt{n_0(n_1+1)}\ket{n_{-1},\,n_0-1,\,n_1+1}\,.
\end{align}
For the case of generic $l$, with a very similar analysis, one can write
\begin{widetext}
\begin{align}
  &\hat{S}^+\ket{n_{-l},\,\ldots,\,n_l}=\sum_{m=-l}^{l-1}\sqrt{l(l+1)-m(m+1)}\sqrt{n_{m}(n_{m+1}+1)}\ket{n_{-l},\,\ldots,\,n_{m}-1,\,n_{m+1}+1,\,\ldots,\,n_l}\,.
\end{align}
\end{widetext}
At this point one can interpret the state $\ket{n_{-l},\,\ldots,\,n_l}$ as the tensor product of bosonic modes with occupation number $n_m$. Introducing the bosonic creation and destruction operators for these bosonic modes $\opbdag{m}$, $\opb{m}$ with $[\opb{m},\,\opbdag{m'}]=\delta_{m\,m'}$ one can immediately write
\begin{widetext}
\begin{align}
  &\hat{S}^+\ket{n_{-l},\,\ldots,\,n_l}=\sum_{m=-l}^{l-1}\sqrt{l(l+1)-m(m+1)}\,\opbdag{m+1}\opb{m}\ket{n_{-l,\,}\ldots,\,n_{m},\,n_m,\,\ldots,\,n_l}\,.
\end{align}
\end{widetext}
The $n_l$ are promoted to operators $\hat{n}_m=\opbdag{m}\opb{m}$, and obey the constraint $\sum_{m=-l}^l\hat{n}_m=N$, as we have seen before. In conclusion, inside the fully symmetric Hilbert subspace generated by the states (Eq.~\eqref{defio:eqn} for generic $l$)
\begin{equation}
  \ket{n_{-l}\,\ldots\,n_l}\equiv\frac{1}{\sqrt{N!\,\prod_m n_{m}!}}\hat{P}\ket{\underbrace{(-l\ldots-l)}_{n_{-l}}\ldots\underbrace{(l\ldots l)}_{n_{l}}}\,,
\end{equation}
one has the mapping
\begin{equation}
  \hat{S}^+=\sum_{m=-l}^{l-1}\sqrt{l(l+1)-m(m+1)}\opbdag{m+1}\opb{m}\,.
\end{equation}
Using similar arguments one can prove that
\begin{align}
  \hat{S}^-&=\sum_{m=-l}^{l-1}\sqrt{l(l+1)-m(m-1)}\opbdag{m}\opb{m+1}\nonumber\\
  \hat{S}^z&=\sum_{m=-l}^l\hat{n}_m\,,
\end{align}
where $\hat{S}^z\equiv\sum_j\hat{s}_j^z$, $\hat{S}^-\equiv\sum_j\hat{s}_j^-$.
\section{Different estimate of the decay time in DTWA} \label{app2:sec}
\begin{figure}
	\begin{tabular}{c}
		\begin{overpic}[width=58mm]{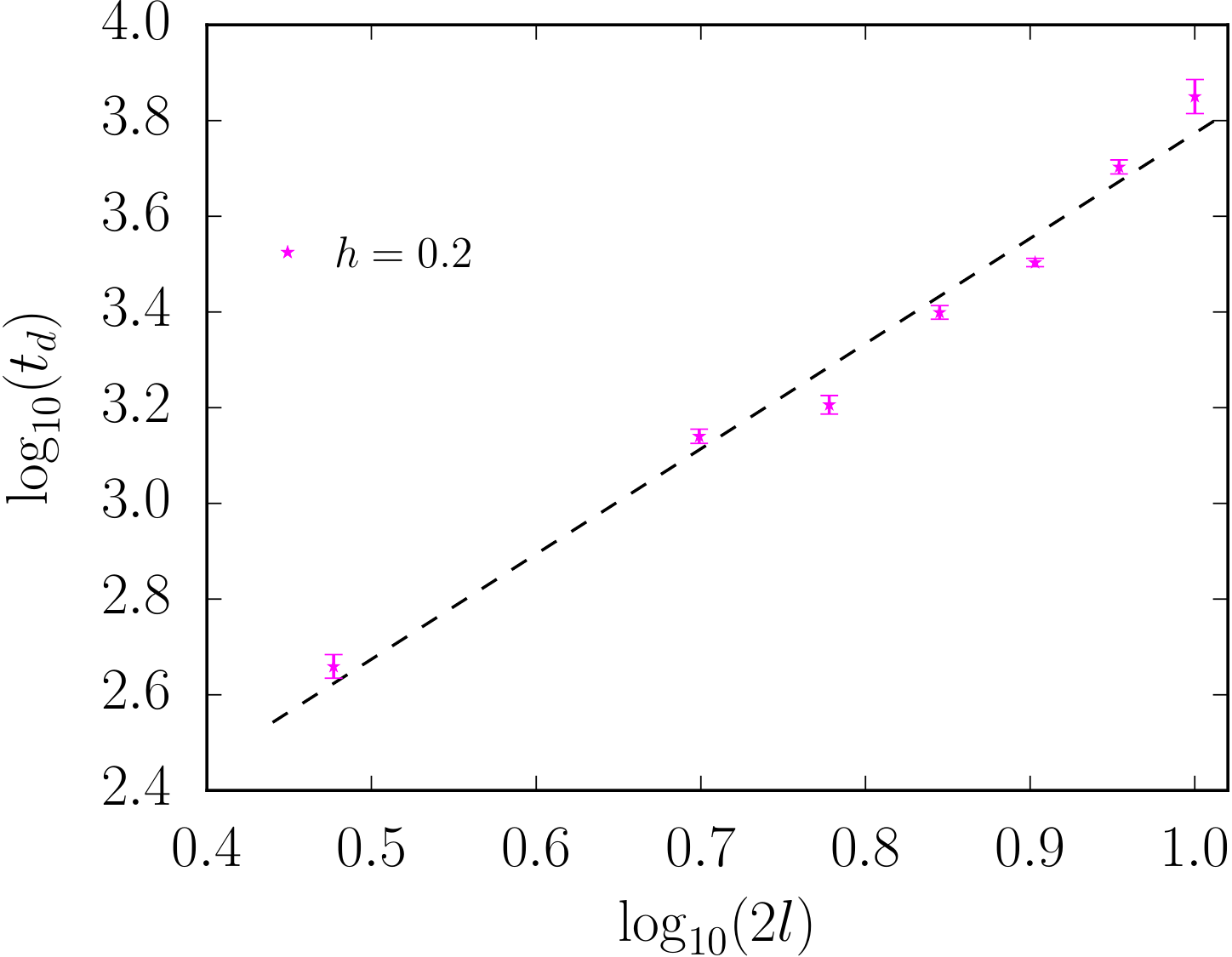}\put(-1,69){(a)}\end{overpic}\\
		\begin{overpic}[width=60mm]{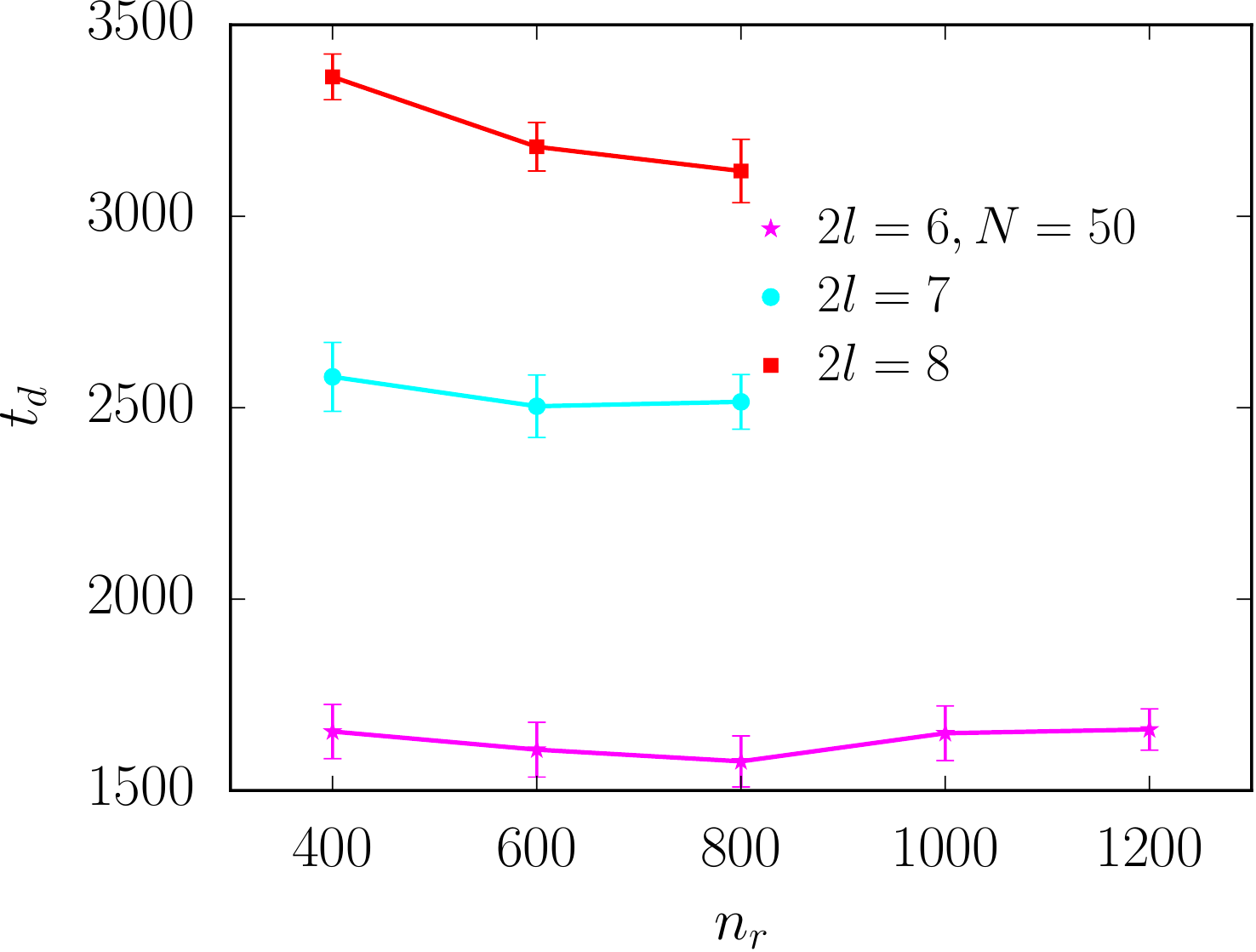}\put(-1,69){(b)}\end{overpic}
	\end{tabular}
	\caption{(a) Decay time of the system as a function of $L_{bin}$. (b) Decay time of the system as a function of $n_r$. Numerical parameters: $N=50, h=0.2, \phi=\pi$, $K=0.3, \tau=0.6, $ and $n_r=800$.} 
	\label{fig6}
\end{figure}

The exponential decay found in Fig.~\ref{fig5}(a) does not last forever and at some point the period-doubling order parameter starts oscillating around 0, as we have seen in Fig.~\ref{fig2}(a). Let us call $t^*$ the first value of the stroboscopic time where the order parameter vanishes. We define
\begin{equation}
t_d=\tau\frac{\sum_{n=1}^{n_{max}}n\mathcal{O}(n\tau)}{\sum_{n=1}^{t^*/\tau}\mathcal{O}(n\tau)}~.
\end{equation}
As we can see in Fig.~\ref{fig6}(a), $t_d$ increases with $l$ as a power law, $t_d\sim l^\gamma$. From a linear fit of the bilogarithmic plot we find $\gamma\simeq 2.20$, in perfect agreement with the finding of Sec.~\ref{reso:sec}. The errorbars for $t_d$ come from the errorbars for $t^*$. We evaluate the latter from the errorbar in $O(n\tau)$ (evaluated as described in Sec.~\ref{dtwa:sec}) which gives rise to an error in the time of first vanishing $t^*$. In Fig.~\ref{fig6}(b) we plot $t_d$ versus $n_r$. In the cases where we can numerically afford $n_r>800$, we see that for $n_r=800$ we have already attained convergence. For larger values of $l$ we cannot go beyond that value, but the clear scaling with $l$ suggests that a satisfying convergence has been already attained for this value of $n_r$.
%

\begin{thebibliography}{59}
\bibitem [{not({\natexlab{b}})}]{notello}%
  \BibitemOpen
  \href@noop {} {} \bibinfo {note} {Also the evolution
  across a kick is immediate. It is easy to see that one must integrate in the
  interval $[0,1]$ the equations $i\frac{\ud}{\ud
  t}\beta_m(t)=\left\{\frac{\phi}{2}-\frac{K}{2l}\left[\sum_{m'=-l}^{l-1}\mathcal{S}(l,m')\Real\left(\beta_{m'}^*(t)\beta_{m'+1}(t)\right)\right]\right\}\nonumber\\
  \cdot[\mathcal{S}(l,m)\,\beta_{m+1}(t)+\mathcal{S}(l,m-1)\,\beta_{m-1}(t)]$,
  with $\mathcal{S}(l,m)\equiv\sqrt{l(l+1)-m(m+1)}$, taking as initial
  conditions the ones immediately before the kick.}\BibitemShut {Stop}%
\end{thebibliography}
%

\end{document}